\documentclass[12points]{nature}
\usepackage{abstract}
\usepackage[utf8]{inputenc}
\usepackage{color}
\usepackage{xcolor}
\usepackage{tabularx}
\usepackage{rotating}
\usepackage[normalem]{ulem}
\usepackage{hyperref}
\usepackage{amsmath,amssymb}
\usepackage{multibib}
\newcites{main}{Main document}%
\usepackage{caption}

\usepackage{graphicx}
 \makeatletter
 \let\saved@includegraphics\includegraphics
 \AtBeginDocument{\let\includegraphics\saved@includegraphics}
 
 \makeatother

\newcommand{\spacingNew}[1]{\renewcommand{\baselinestretch}{#1}\large\normalsize}

\newcommand {\xmm} {\textsl{XMM-Newton}}

\newcommand {\nustar} {\textsl{NuSTAR}}

\newcommand {\nicer} {\textsl{NICER}}

\def \rsun {\ifmmode$R$_{\odot}\else R$_{\odot}$}

          \font\sixrm=cmr6

\def\teq#1{$\, #1\,$}                         
\def\dover#1#2{\hbox{${{\displaystyle#1 \vphantom{(} }\over{
   \displaystyle #2 \vphantom{(} }}$}}

\def \hcm {\hbox {\ifmmode $ atoms cm$^{-2}\else atoms cm$^{-2}$\fi}}

\def\approxgt{\mathrel{\hbox{\rlap{\lower.55ex \hbox {$\sim$}}
        \kern-.3em \raise.4ex \hbox{$>$}}}}
\def\approxlt{\mathrel{\hbox{\rlap{\lower.55ex \hbox {$\sim$}}
        \kern-.3em \raise.4ex \hbox{$<$}}}}

\def \arcsec {\hbox{$^{\prime\prime}$}}

\newcommand\T{\rule{0pt}{2.6ex}}
\newcommand\B{\rule[-1.2ex]{0pt}{0pt}}

\def \src {SGR~1935+2154}

          \font\sixrm=cmr6       

\def\sigt{\sigma_{\hbox{\sixrm T}}}
\def\taut{\tau_{\hbox{\sixrm T}}}

\title{{\bf Magnetar spin-down glitch clearing the way for FRB-like bursts and a pulsed radio episode}}

\author{
G.~Younes$^{1,2}$, M.~G.~Baring$^{3}$, A.~K.~Harding$^{4}$, T.~Enoto$^{5}$, Z.~Wadiasingh$^{1,6,7}$, A.~B.~Pearlman$^{8,9,10}$, W.~C.~G.~Ho$^{11}$, S.~Guillot$^{12, 13}$, Z.~Arzoumanian$^{1}$, A.~Borghese$^{14,15}$, K.~Gendreau$^{1}$, E.~G\"{o}\u{g}\"{u}\c{s}$^{16}$,  T.~G\"uver$^{17}$, A.~J.~van~der~Horst$^{2}$, C.-P.~Hu$^{18}$, G.~K.~Jaisawal$^{19}$, C.~Kouveliotou$^{2}$, L.~Lin$^{20}$, W.~A.~Majid$^{21,10}$}

\begin{document}

\maketitle
\begin{itemize}

\item[1]{Astrophysics Science Division, NASA/GSFC, Greenbelt, MD 20771, USA}

\item[2]{Department of Physics, The George Washington University, 725 21st St. NW, Washington, DC 20052, USA}

\item[3]{Department of Physics and Astronomy - MS 108,
Rice University, 6100 Main Street, Houston, Texas 77251-1892, USA}

\item[4]{Theoretical Division, Los Alamos National Laboratory, Los Alamos, NM 87545, USA}

\item[5]{Extreme Natural Phenomena RIKEN Hakubi Research Team, Cluster for Pioneering Research, RIKEN, Wako, Saitama 351-0198, Japan}

\item[6]{Department of Astronomy, University of Maryland, College Park, Maryland 20742, USA}
\item[7]{Center for Research and Exploration in Space Science and Technology, NASA/GSFC, Greenbelt, Maryland 20771, USA}

\item[8]{Department of Physics, McGill University, 3600 rue University, Montréal, QC H3A 2T8, Canada}
\item[9]{McGill Space Institute, McGill University, 3550 rue University, Montréal, QC H3A 2A7, Canada}
\item[10]{Division of Physics, Mathematics, and Astronomy, California Institute of Technology, Pasadena, CA 91125, USA}

\item[11]{Department of Physics and Astronomy, Haverford College, 370 Lancaster Avenue, Haverford, PA 19041, USA}

\item[12]{IRAP, CNRS, 9 avenue du Colonel Roche, BP 44346, F-31028 Toulouse Cedex 4, France}
\item[13]{Universit\'{e} de Toulouse, CNES, UPS-OMP, F-31028 Toulouse, France}

\item[14]{Institute of Space Sciences (ICE, CSIC), Campus UAB, Carrer de Can Magrans s/n, E-08193, Barcelona, Spain}
\item[15]{Institut d’Estudis Espacials de Catalunya (IEEC), Carrer Gran Capit. 2–4, E-08034 Barcelona, Spain}

\item[16]{Sabanc{\i} University, Faculty of Engineering and Natural Sciences, 34956, \.Istanbul, Turkey}

\item[17]{Istanbul University, Science Faculty, Department of Astronomy and Space Sciences, Beyaz\i t, 34119, Istanbul, Turkey}

\item[18]{Department of Physics, National Changhua University of Education, Changhua, 50007, Taiwan}

\item[19]{National Space Institute, Technical University of Denmark, Elektrovej 327-328, DK-2800 Lyngby, Denmark}

\item[20]{Department of Astronomy, Beijing Normal University, Beijing 100875, People's Republic of China}

\item[21]{Jet Propulsion Laboratory, California Institute of Technology, Pasadena, CA 91109, USA}

\end{itemize}


\begin{abstract}

\noindent{\bf Magnetars are a special subset of the isolated neutron star family, with X-ray and radio emission mainly powered by the decay of their immense magnetic fields.  Many attributes of magnetars remain poorly understood: spin-down glitches or the sudden reductions in the star's angular momentum, radio bursts reminiscent of extra-galactic Fast Radio Bursts (FRBs), and transient pulsed radio emission lasting months to years. Here we unveil the detection of a large spin-down glitch event ($|\Delta\nu/\nu| = 5.8_{-1.6}^{+2.6}\times10^{-6}$) from the magnetar SGR~1935+2154 on 2020 October 5 (+/- 1 day). We find no change to the source persistent surface thermal or magnetospheric X-ray behavior, nor is there evidence of strong X-ray bursting activity. Yet, in the subsequent days, the magnetar emitted three FRB-like radio bursts followed by a month long episode of pulsed radio emission. Given the rarity of spin-down glitches and radio signals from magnetars, their approximate synchronicity suggests an association, providing pivotal clues to their origin and triggering mechanisms, with ramifications to the broader magnetar and FRB populations. We postulate that impulsive crustal plasma shedding close to the magnetic pole generates a wind that combs out magnetic field lines, rapidly reducing the star's angular momentum, while temporarily altering the magnetospheric field geometry to permit the pair creation needed to precipitate radio emission.}
\end{abstract}


\newpage

\src\ is an isolated neutron star displaying hot and luminous soft X-ray emission pulsed\cite{Israel-2016-MNRAS} at a spin frequency $\nu\approx0.308$~Hz while slowing down at a nominal rate of about $-1.4 \times 10^{-12}$~Hz~s$^{-1}$. If attributed to magnetic dipole braking, these spin properties imply a dipole magnetic field strength $B\approx2.2\times10^{14}$~G at the equator and a young spin-down age $\tau\approx3.6$~kyr. \src\ is also a prolific burster, capable of displaying in a matter of minutes hundreds of bright millisecond-duration X-ray bursts\cite{younes20ApJ1935}, with luminosities exceeding $1.0\times10^{41}$~erg~s$^{-1}$. Hence, \src\ belongs to the small, special group of isolated neutron stars known as magnetars, for which the very strong magnetic field powers their many emission characteristics. Due to their extreme variable nature and large magnetic energy budget, magnetars are the leading suspect for the sources of enigmatic bright millisecond radio flashes of extragalactic origin known as Fast Radio Bursts (FRBs)\cite{petroff2019AARv}. Indeed, in a rare occurrence to date, on 2020 April 28 during a period of intense X-ray bursting activity\cite{younes20ApJ1935}, \src\ emitted a radio burst with a luminosity approaching extragalactic FRBs. This discovery provided the first evidence for the nature of the progenitor of at least some FRBs \cite{Bochenek20:1935,chime2020:1935}.

Since this event, we have been monitoring \src\ regularly with several X-ray instruments, most notably in the soft, 1--3~keV band with the \nicer\ and \xmm\ telescopes. In this band, X-rays from the source are dominated by the pulsed, surface thermal emission, enabling us to track the evolution of its spin ephemerides. During a particularly heavy cadence observational period covering the October 1st to November 27th dates, we were able to employ a phase-coherent timing analysis, i.e., tracking the time-of-arrival (TOA) of X-ray pulses from the source with a precise timing model. The pulse arrival time of \src\ from October 6th to November 27th is well predicted, with an accuracy that is a few percent of the source spin-period, from a simple timing model that includes the frequency and its first and second derivatives. However, this model fails to predict the pulse arrival time from the 1st and the 2nd of October, showing an offset of about half a rotation just 3.5 to 5 days later. Attempting to model these residuals with the inclusion of higher order frequency derivatives fails to provide a statistically acceptable fit (See methods, Figure~\ref{fig:timModF2F3}).

\begin{figure*}[h!]
\begin{center}
  \includegraphics[angle=0,width=0.71\textwidth]{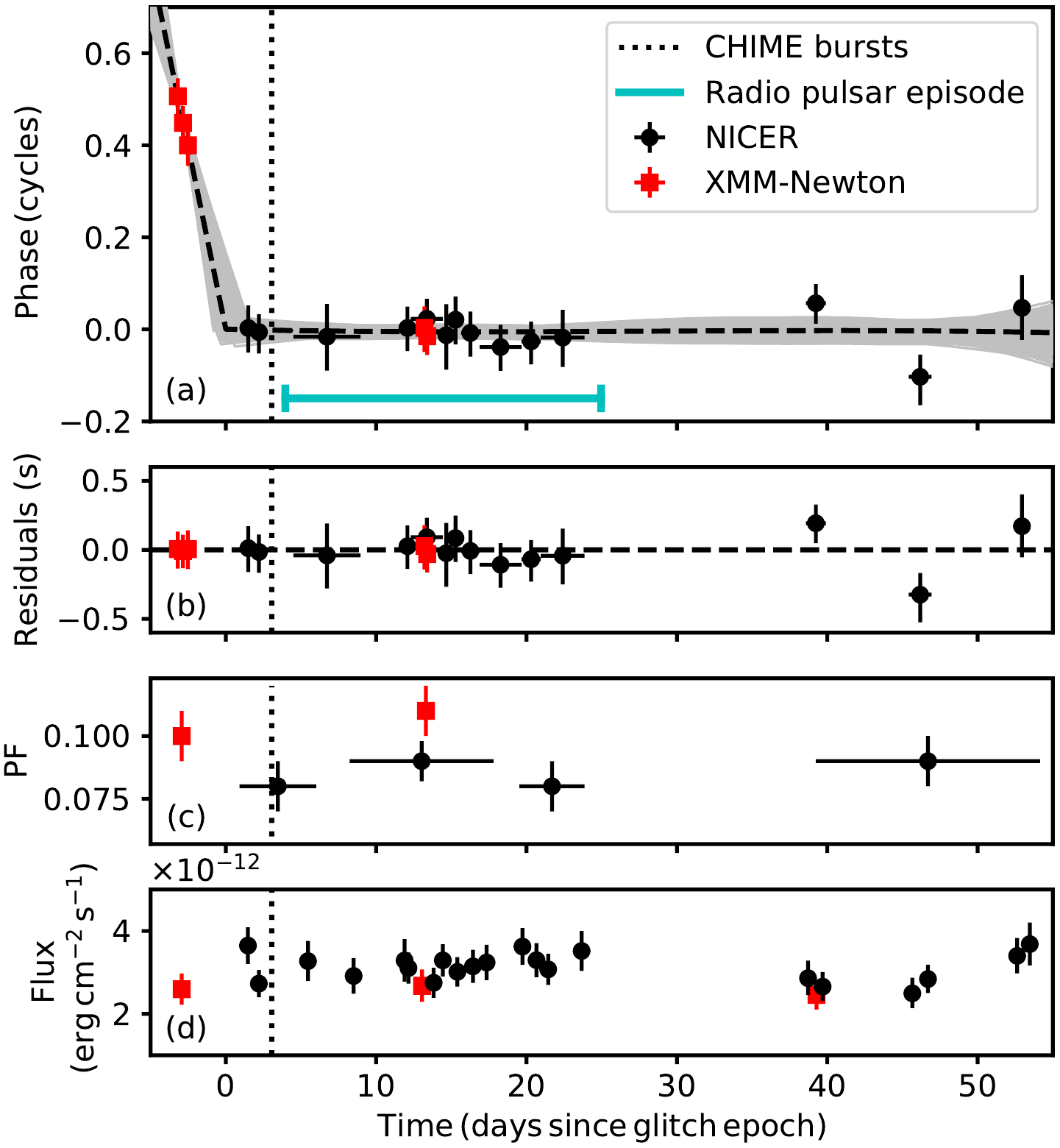}
  \spacingNew{1}
\caption{{\sl Panel (a).} Phase residuals, in rotational cycles, of the SGR~1935+2154 X-ray pulses, according to the best-fit timing model that excludes the three earliest data-points. Dashed line is the best-fit timing model to the full baseline, which includes a spin-down glitch (a sudden change to the spin frequency), along with its $1\sigma$ uncertainty (gray lines). The cyan horizontal line marks the time period during which FAST detected radio pulsations (Zhu et al. in prep., private communication). Black dots and red squares represent the pulse-phase of \nicer\ and \xmm\ data, respectively. The horizontal line on each indicates the temporal extent that encompassed enough exposure so that a pulse can be measured. {\sl Panel (b)}. Pulse residuals in seconds from the best-fit timing model, including the glitch, to the full baseline. {\sl Panel (c).} The root-mean-square pulsed fraction of the X-ray emission in the energy range 1-3 keV derived by combining several individual observations, as shown with the horizontal bar. {\sl Panel (d).} The 1-10 keV absorption-corrected flux from individual \nicer\ and \xmm\ observations. In all panels, the vertical dotted line is the occurrence of the CHIME bursts on 2020 October 08. The vertical lines on each data point are the $1\sigma$ uncertainty on those measurements. The \xmm\ observation around day 40 is considered in tandem with the nearby \nicer\ data when performing the timing analysis.}
\label{fig:antGli}
\end{center}
\end{figure*}

The sharp and large pulse-phase offset observed in the early October data is reminiscent of the glitching behavior observed in pulsars and magnetars when they exhibit a sudden jump in spin-frequency (i.e., $\Delta\nu$) at a well-defined epoch $t_{\rm g}$. Indeed, a timing model that includes a glitch provides an accurate prediction of the pulse TOA for the full October and November time period (Figure~\ref{fig:antGli} and Table\ref{tabTimSol}). In this model, we find that a frequency jump $\Delta\nu = 1.8_{-0.5}^{+0.7} \times 10^{-6}$~Hz (corresponding to a fractional change $\Delta\nu/\nu = 5.8_{-1.6}^{+2.6} \times 10^{-6}$) occurred at a glitch epoch $t_{\rm g}=59127.2_{-0.7}^{+1.0}$~MJD or October 5th. We note that the positive frequency jump is required to explain the early TOAs relative to our reference epoch (59141.0~MJD), implying that the source experienced a negative $\Delta\nu$ frequency jump at $t_{\rm g}$. The corresponding loss of the magnetar rotational kinetic energy due to this abrupt spin-down event is of the order of $3.0\times10^{40}$~erg. This phenomenon of a spin-down glitch, also referred to as "anti-glitch", has been conclusively observed from one other magnetar, 1E~2259+586\cite{Archibald-2013-Nature,younes20ApJ:2259}. The spin-down glitch magnitude, as well as the fractional change in the case of \src, are about one order of magnitude larger compared to the three spin-down glitches observed so far from 1E~2259+586 over a period of 20 years of observations\cite{dib14ApJ,younes20ApJ:2259}. Other potential spin-down glitches have been reported from other magnetars, most notably is the case of SGR~1900+14 where a spin-down event, an order of magnitude larger than in \src, occurred during an 80~day gap around the time of its August 1998 giant flare\cite{woods99ApJ:1900}.

We searched for X-ray variability in the properties of \src\ associated with the spin-down glitch epoch, but found none. For instance, the soft thermal and hard non-thermal X-ray flux throughout the October period remained at a constant level, as did the surface temperature and the non-thermal spectral shape (Figures~\ref{fig:antGli}, \ref{fig:specXMM}, and Table~\ref{specParam}). Moreover, the broad complex pulse profile shape as well as the pulsed fraction remained stable throughout the same period (Figure~\ref{fig:timEvol}). Finally, we detected no magnetar-like short bursts from \src\ in any of our X-ray observations, in line with the lack of detection of strong bursting activity by large field-of-view hard X-ray instruments (e.g., Fermi/GBM, Swift/BAT). Any variability in the source intrinsic X-ray flux associated with the glitch is either constrained to a $3\sigma$ upper-limit of about $10^{-12}$~erg~s$^{-1}$~cm$^{-2}$ based on a \nustar\ observation which occurred on 2020 October 4 (Table~\ref{specParam}), or confined to a 1-day interval between MJD 59127.18, the end of the latter observation, and 59128.06, the start of a \nicer\ observation on October 6 (see methods, Figure~\ref{fig:nusLC}).

While the spin-down glitch is apparently X-ray silent, \src\ exhibited profound changes at radio frequencies. Firstly, three moderately bright FRB-like radio bursts were detected 3 days following the glitch epoch\cite{good20ATel14074}. These bursts have a duration of about a few milliseconds each and occurred during a single rotational period of the source. These properties resemble those of the previous radio bursts detected from \src, including the April 28 event\cite{chime2020:1935,kirsten2020} (albeit much fainter). Less than one day following the radio-burst detections, the FAST radio telescope observed \src\ and detected, for the first time, the emergence of a pulsed radio component\cite{zhu20ATel14084}. This component was not detected in any of the numerous previous radio observations of the source\cite{lin2020Natur}, including as recent as August 28 with the FAST radio dish.

The spin-down glitch may constitute a sudden transfer of angular momentum ${\cal L}$ away from the star, nominally carried by a particle wind along open field lines. This very likely originates from the surface of the star, yet it could be coupled to a release of magnetic energy stored in twisted magnetospheric field configurations\cite{Parfrey-2012-ApJ}. Other hypotheses for the origin of abrupt spin-down events exist. A sudden increase in the oblateness and/or moment of inertia $I$ was a hypothesis used to explain the first anti-glitch observed for 1E 2259+586 \cite{Mastrano-2015-MNRAS}. Alternatively, angular momentum transfer to a more slowly spinning inner crust could be a seed for strong spin down \cite{Thompson-2000-ApJ}. Yet, the contemporaneous detection of the spin-down glitch with the FRB-like bursts and the radio-pulsar episode appears remarkable, and suggests a causal connection given the rarity of each phenomenon (see methods). This is strongly suggestive of an external process, and so here we explore the wind scenario and its implications.

An ephemeral, strong wind emanating from the surface and passing through the magnetosphere on opened field lines will naturally generate strong angular momentum loss; such winds have been invoked to address general plasma loading of magnetar magnetospheres\cite{Harding-1999-ApJ}. Using conservation of the total angular momentum ${\cal L}$, for a magnetar of mass $M$, one deduces that the cumulative mass $\delta m$ deposited in the wind satisfies $\delta m/M \sim 10^{-10}$ and a luminosity/mass loss rate of $L_{\rm w}\sim 7 \times 10^{39}$~erg~s$^{-1}$, under the assumption of a dipole near the surface and a putative transient wind duration of 10 hours (see Methods). This mass loss should be considered an upper-limit and is likely 2 to 3 orders of magnitude lower if the field configuration is strongly twisted near the poles, corresponding to a larger open field-line polar cap that naturally arises during abrupt mass-shedding \cite{Harding-1999-ApJ}. In such a case, radiative efficiencies of a few percent or less would then yield X-ray luminosities low enough to be consistent with the non-detection of any transient flaring activity associated with the glitch epoch (see Methods).

The extremely high opacity conditions during the strong wind phase generally preclude the establishment of electric potential gaps and associated electron acceleration, and subsequent curvature radiation or resonant Compton upscattering, and electron-positron pair creation. These are all elements long deemed essential to radio emission in pulsars\cite{Sturrock-1971-ApJ}, and most are likely for the generation of fast radio bursts\cite{Wadiasingh-2020-ApJ}. Accordingly, one does not expect radio emission during the glitch. But what conditions prevail after this abrupt spin-down event that might permit FRB-like emission as well as the pulsed radio signal? The answer may lie in ephemeral modifications to the magnetic field geometry.

Powerful winds are well-known to comb out magnetospheric fields to become almost radial\cite{Bai-2010-ApJ,Kalapotharakos-2012-ApJ}. The large mass loss of the strong wind implies that it probably originates at the stellar surface, likely connected to sub-surface structural rearrangements in the outer crust and its embedded fields. Such alterations may seed a temporary perturbation to the magnetospheric field morphology that may enhance conditions for pair creation and radio emission. Specifically, if the magnetic field curvature includes evolving toroidal (twisted) components that are vestiges of the powerful wind phase, the pair creation rate can be increased substantially relative to that for photon splitting (see Methods). This prospect is underpinned by the extreme sensitivity of the pair creation rate to the magnetic field strength, the field line curvature, and the directional beaming of radiation\cite{Harding-1978-ApJ,Baring-2001-ApJ}. As the wind further abates, the twisted and curved field lines re-establish themselves at their pre-glitch configuration, so that photon splitting again becomes more potent in suppressing pair creation\cite{Baring-1998-ApJL}, likely shutting down the radio signal.

Such field geometry adjustments cannot be sufficient to modify the surface thermal emission below 4 keV (e.g., through particle bombardment\cite{Beloborodov-2016-ApJ}) and the hard X-ray tail signal above $\sim 4$ keV, since these are not impacted by the spin-down glitch, see Fig.~\ref{fig:specXMM}. Yet this change may be confined to polar locales and just enough to permit the triggering of radio bursts\cite{Wadiasingh-2020-ApJ,metzger19MNRAS} and pulsed radio emission. These may accompany a residual wind phase at a modest level commensurate with the long-term spin-down rate, with relaxation due to Ohmic dissipation back to the long-term field configuration in concert with the radio turn off a month later.  In essence, perhaps the spin-down glitch temporarily moves the radio ``death line'' on the pulsar period-period derivative diagram\cite{Ruderman-1975-ApJ} to longer periods due to its field geometry modifications. The persistent surface X-ray signal emanating from a considerable range of closed field line colatitudes just continues unaltered and unabated. Our results highlight the necessity for deeper theoretical studies of field morphology associated with magnetospheric plasma loading and its evolution via Ohmic dissipation, in concert with gamma-ray opacity and pair creation considerations to unravel the physical conditions and mechanisms responsible for FRB-like bursts and pulsed radio emission in magnetars.

\newpage 

\begin{table}
\rmfamily
\captionsetup{justification=centering}
\caption{Best fit spin parameters for the 2020 October 1 to November 27 period.}
\label{tabTimSol}
\begin{center}
\resizebox{0.5\textwidth}{!}{
\begin{tabular}{l r}
\hline
\hline
R.A. (J2000) \T\B & 19:35:41.64 \\ 
Decl. (J2000) \T\B & 21:54:16.9 \\ 
Time Scale \T\B & TDB \\
Ephemeris \T\B & DE405\\
Epoch (MJD) \T\B & 59141.0 \\
$\nu$ (Hz) \T\B & $0.30789626(2)$ \\
$\dot{\nu}$ (Hz s$^{-1}$) \T\B &  $-3.52(3)\times10^{-12}$ \\
$\ddot{\nu}$ (Hz s$^{-1}$) \T\B &  $1.9(3)\times10^{-19}$ \\
$t_{\rm g}$ (MJD, TDB) \T\B & $59127.2^{+1.0}_{-0.7}$ \\
$\Delta\nu$ (Hz) \T\B &  $1.8_{-0.5}^{+0.7}\times10^{-6}$  \\
Valid Range (MJD) \T\B &  59123.7--59180.5 \\
$\chi^2$/dof  &  15/13 \\
RMS residual (ms) \T\B &  105 \\
\hline
$\Delta\nu/\nu$ \T\B & $5.8_{-1.6}^{+2.6}\times10^{-6}$ \\
\hline
\end{tabular}}
\end{center}
\end{table}

\pagebreak
\newpage 

\section*{References}
\bibliographystyle{naturemag}
\bibliography{bibMain.bib}

\newpage 

\section*{Data Availability}
NICER raw data and calibrated level-2 data files were generated at the Goddard Space Flight Center large-scale facility. These data files are publicly available and can be found \href{https://heasarc.gsfc.nasa.gov/FTP/nicer/data/obs/}{at this link}. \xmm\ and \nustar\ data files are also publicly available from the \href{http://nxsa.esac.esa.int/nxsa-web/}{\xmm\ Science archive} and the \href{https://heasarc.gsfc.nasa.gov/W3Browse/all/numaster.html}{NUMASTER} table. Level 3 data supporting the findings of this study are available from the corresponding authors upon request. 

\section*{Code Availability}
Reduction and analysis of the data were conducted using publicly available  codes  provided  by  the  High  Energy  Astrophysics Science Archive Research Center (HEASARC), which is a service of the Astrophysics Science Division at NASA/GSFC and the High Energy Astrophysics Division of the Smithsonian Astrophysical Observatory. For \nicer\ and \nustar, we used NICERDAS version v008c and NUSTARDAS version v2.1.1, respectively, part  of  HEASOFT  6.29c (https://heasarc.gsfc.nasa.gov/docs/software/lheasoft). For \xmm\ we utilize the publicly available Science Analysis Software (SAS) version 19.1.0. Spectral analysis was conducted using Xspec version 12.12.0g (https://heasarc.gsfc.nasa.gov/docs/xanadu/xspec/). The EMCEE MCMC sampler is a public software available \href{https://emcee.readthedocs.io/en/stable/}{here}. Custom codes for the timing analysis routines are available upon reasonable request from the corresponding author.

\section*{Competing Interests}

The authors declare that they have no competing financial interests.

\newpage

\section*{Acknowledgements}

A portion of this work was supported by NASA through the NICER mission and the Astrophysics Explorers Program. This research has made use of data and software provided by the High Energy Astrophysics Science Archive Research Center (HEASARC), which is a service of the Astrophysics Science Division at NASA/GSFC and the High Energy Astrophysics Division of the Smithsonian Astrophysical Observatory. G.~Y. research is supported by an appointment to the NASA Postdoctoral Program at the Goddard Space Flight Center, administered by Oak Ridge Associated Universities under contract with NASA. M.~G.~B. acknowledges the support of the National Science Foundation through grant AST-1813649. A.B.P. is a McGill Space Institute (MSI) Fellow and a Fonds de Recherche du Quebec -- Nature et Technologies (FRQNT) postdoctoral fellow. SG acknowledges the support of the Centre National d'Etudes Spatiales (CNES). T.E. acknowledges Hakubi projects of Kyoto University and RIKEN, and supported by JSPS/ MEXT KAKENHI grant nombers 15H00845 and 17K18776. W.C.G.H. acknowledges support through grant 80NSSC22K0397 from NASA. AB is supported by a Juan de la Cierva fellowship. C.-P.H.~acknowledges support from the Ministry of Science and Technology in Taiwan through grant MOST 109-2112-M-018-009-MY3. W.A.M acknowledges support from the Jet Propulsion Laboratory, California Institute of Technology, under a Research and Technology Development Grant through a contract with the National Aeronautics and Space Administration. U.S. government sponsorship is acknowledged. G.~Y. would like to thank Victoria Kaspi and Tod Strohmayer for providing constructive comments on the manuscript and Paul Ray for guidance on the timing analysis.

\section*{Author contributions statement}

G.Y. performed the data analysis and contributed to the writing of the associated text. M.G.B. led the interpretative elements and was responsible for the writing of the associated text. A.K.H., T.E., Z.W., W.C.G.H., S.G., A.B., A.B.P., T.G., A.J.v.d.H., C.-P.H., E.G., G.K.J., C.K., L.L., and W.A.M. contributed to the discussion and editing of the manuscript. A.B.P. and G.Y. were responsible for acquiring the majority of the NICER data through the Director’s Discretionary Time. E.G. and L.L. were responsible for acquiring part of the \xmm\ and \nustar\ data. A.B. was responsible for acquiring the \xmm\ October 1st data and part of the \nicer\ data. K.C.G. is the NICER principal investigator; he approved the Director’s Discretionary Time observations. Z.A. is the NICER project scientist and deputy principal investigator; he contributed to the scheduling of the NICER observations.

\section*{Correspondence}
Correspondence and requests for materials should be addressed to G.Y.~(email: george.a.younes@nasa.gov) and/or M.G.B. (email: baring@rice.edu)

\newpage

\section*{Methods}
\label{sec:methods}

\noindent {\bf \nicer\ observations and data reduction.}
\nicer\cite{gendreau16SPIE} is a soft X-ray telescope mounted on the International Space Station, sensitive to photon energies in the range 0.3-12~keV. It consists of 56 co-aligned X-ray concentrating optics, of which 52 are currently operational, providing a collecting area of about 1900~cm$^2$ at 1.5~keV. \src\ was observed extensively with \nicer\ following the announcement of the 2020 October 8 CHIME radio bursts. For this paper, we analyze the \nicer\ observation IDs 3020560154-75 and 3655010401-02, covering the period 2020 October 06 to 2020 November 27, during which a detailed spectral and phase-coherent timing analysis is carried out (see below). We note that two observations that occurred on 2020 December 03 and 04 (obs IDs 3020560176 and 3020560177) resulted in a combined exposure of 2~ks after background-cleaning, which is insufficient for any meaningful spectral or temporal analysis, hence, ignored. We also perform detailed timing analysis on heavy-cadence \nicer\ observations covering the 2020 June 18 to August 06 period (observation IDs 3655010302-03 and 3020560120-48). Between these two episodes, however, only sparse and short X-ray observations existed which prevented us from phase-coherently connecting them to the focal epoch of this paper. In addition, subsequent to November, \src\ could not be observed for two months due to its proximity to the Sun. For the observations included in our analysis, we use NICERDAS version v008c to create cleaned and calibrated event files, extract spectra, and build light curves, after applying standard filtering to all observations as described in the \nicer\ Data Analysis Guide\footnote{\href{https://heasarc.gsfc.nasa.gov/docs/nicer/data\_analysis/nicer\_analysis\_guide.html}{NICER data analysis guide.}}. Finally, we estimate the background number counts per \nicer-energy channel utilizing the \texttt{nibackgen3C50} tool, and added a conservative 20\% systematic uncertainty to this estimate\cite{remillard2021:3c50}.

\noindent {\bf \xmm\ observations and data reduction.} \xmm\ is an imaging X-ray satellite with several cameras on board, sensitive to photon energies in the range 0.4-10~keV. For this paper, we analyze the \xmm\ observations taken on 2020 October 1 (observation ID 0871191301), October 18 (observation ID 0872390601), and November 12 (observation ID 0872390701), with background-corrected exposures of 61, 29, and 18 ks, respectively. We focus on the EPIC-pn camera\cite{struder01aa}, which operated in prime full-frame mode for all observations, affording a 73~ms time resolution. We performed the cleaning and filtering of the events using the \xmm\ Science Analysis Software (SAS) version 19.1.0. We applied standard filtering to all observations similarly (e.g., only event patterns 0–12 were accepted and good X-ray events with FLAG=0). Furthermore, we excluded intervals of high background flaring activity, e.g., due to solar flares, as measured from source-free full field-of-view light curve. Finally, we extracted source events from a circle centered at the best-fit PSF location as obtained with the SAS task \texttt{eregionanalyse}, having a radius of 60\arcsec, encapsulating 90\% of a point source point spread function. Background events are extracted from a source-free annulus centered at the source with inner and outer radii of 120\arcsec\ and 200\arcsec, respectively. We generated response matrix and ancillary files using the SAS tasks \texttt{rmfgen} and \texttt{arfgen}, respectively. 

\noindent {\bf \nustar\ observations and data reduction.} The focusing hard X-ray telescope NuSTAR (Nuclear Spectroscopic Telescope ARray\cite{harrison13ApJ:NuSTAR}) consists of two identical modules, FPMA and FPMB, sensitive to photon energies in the range 3-79~keV. \nustar\ observed \src\ on 2020 October 4 (obs ID 80602313008), 14 (obs ID 90602332002), and 16 (obs ID 90602332004) with exposures of 40, 20, and 18~ks, respectively. We reduced the NuSTAR data using NuSTARDAS software version 2.1.1 as part of HEASoft 6.29c along with the calibration files version 20201130. We extract source events, light curves, and spectra from a circular region with a 45\arcsec-radius around the source central brightest pixel. We estimate the background contribution to the source from an annulus centered on the source, with an inner and outer radii of 120\arcsec\ and 200\arcsec, respectively.

\noindent {\bf Timing analysis.} We converted all cleaned events time stamps to the Barycentric Dynamical Time (TDB) which measures the photon arrival times at the solar system barycenter. For this purpose, we utilized the JPL ephemerides DE405, and the best known source location as measured with the Hubble Space Telescope\cite{levan18ApJ:1935}. The source small pulsed fraction of around 8\%, and its relative faintness proved problematic to maintaining phase-coherence throughout the extent of the 2020 observations. Nevertheless, the heavy X-ray cadence and deep observations throughout the months of October and November (October 1 to November 27) allowed us to follow the pulse time of arrival to high accuracy. Firstly, we relied on the \xmm\ and \nicer\ observations of October 18 and 19, with an exposure totalling 72~ks to establish an accurate spin-frequency of the source. Using a Z$_{\rm n}^2$ test\cite{buccheri83AApulse} with number of harmonics ${\rm n}=2$, we find the strongest signal in the energy range 1-3~keV at a frequency $\nu=0.3078961(4)$~Hz at the epoch $T_0=59141.0$~MJD (TDB). Using these ephemerides, we measured the pulse arrival time for segments of data from October 1 to November 27 containing approximately 6500 events; the number of events required to detect the pulsed emission at $\sim4.5\sigma$. We employed a non-binned maximum likelihood technique to measure the pulse time of arrival\cite{ray11ApJS,livingstone09ApJ}. We relied on the above high S/N pulse profile to build a model of the pulse shape, consisting of the sum of the first two harmonics of a Fourier series. We then fit this model to each unbinned data segment allowing for a phase-shift $\Delta\phi$. The $1\sigma$ uncertainty on the phase shift was established by using the MCMC sampler \texttt{emcee}\cite{emcee2013PASP}. We assumed a flat prior $\Delta\phi \in [0,2\pi)$ and evolved 32 walkers for 1000 steps.

\begin{figure*}[h!]
\begin{center}
\includegraphics[angle=0,width=0.48\textwidth]{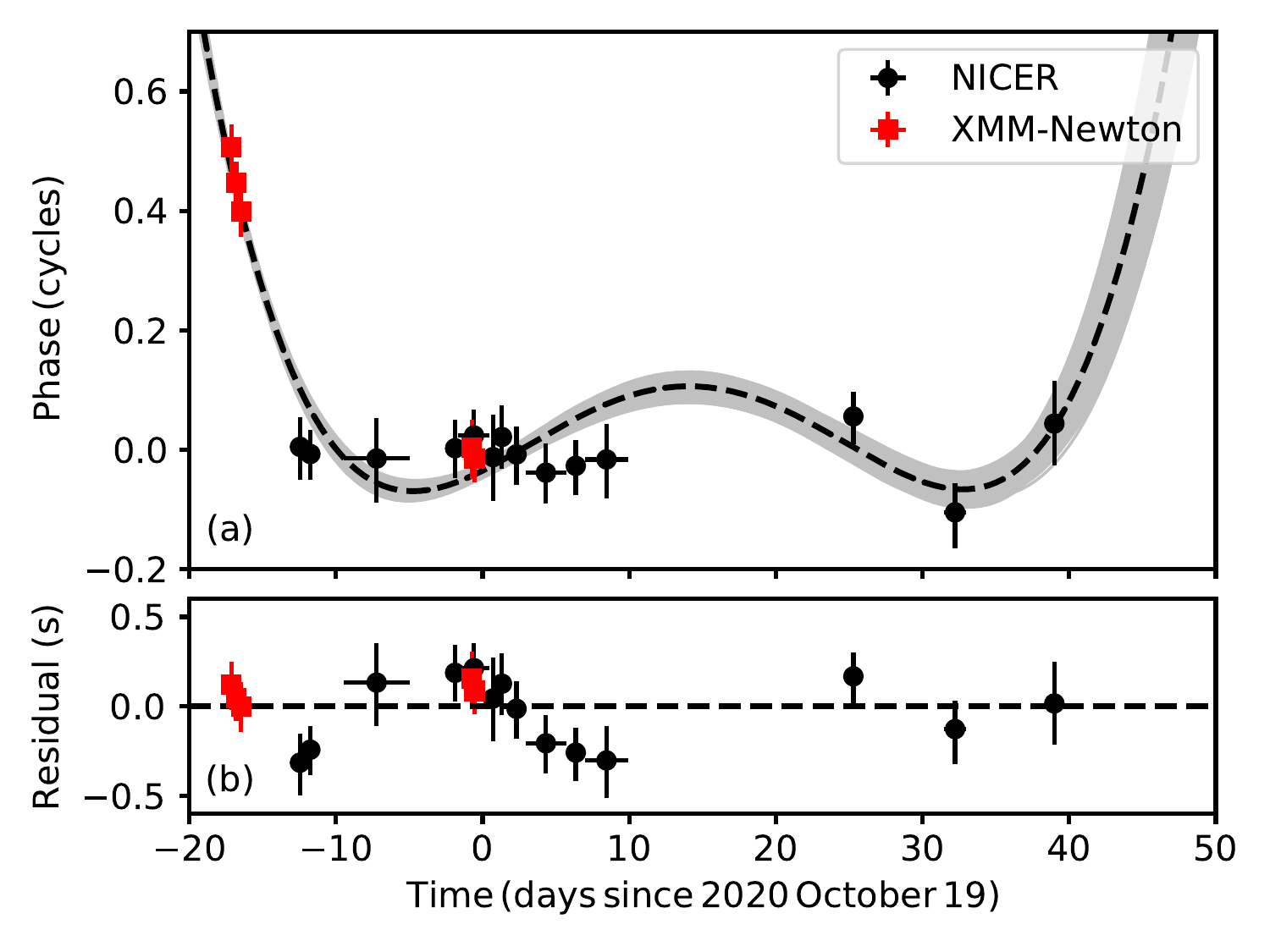}
\includegraphics[angle=0,width=0.48\textwidth]{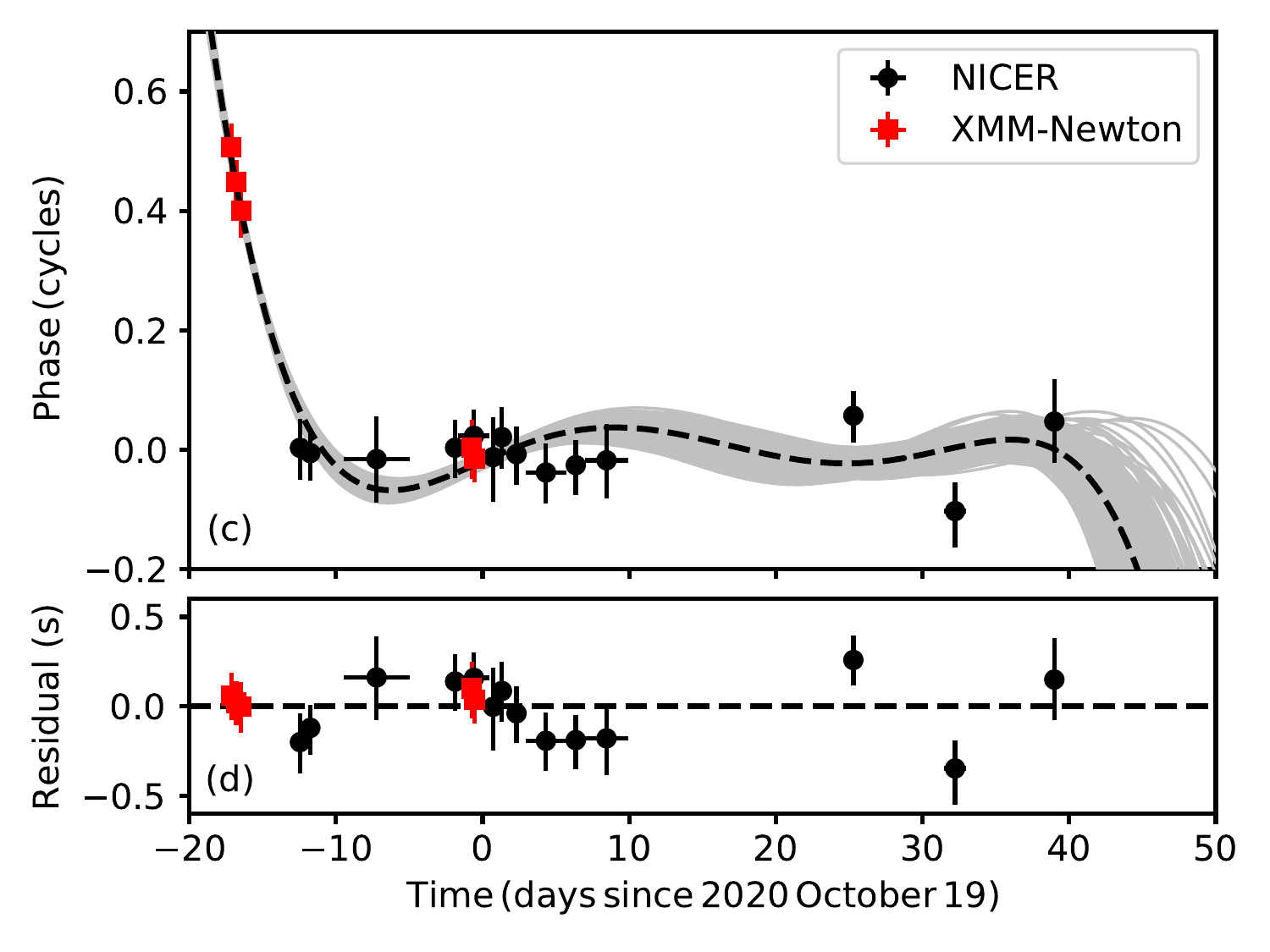}
\spacingNew{1}
\caption{{\sl Panel (a)}. Phase residuals, in rotational cycles, of the SGR~1935+2154 X-ray pulses, according to the best-fit timing model that excludes the three earliest data-points. Black dots and red squares represent the pulse-phases of \nicer\ and \xmm\ data, respectively. The black-dashed line is the best fit smooth timing model to all data points, according to equation~\ref{eqTimMod} and including up to $\dddot{\nu}$. {\sl Panel (b)}. Residuals of the X-ray pulse arrival times in seconds from the latter. {\sl Panel (c)}. Same as (a) except the timing model includes up to $\ddddot{\nu}$. Note that this model has the same number of free parameter as the one shown in Figure~\ref{fig:antGli}. {\sl Panel (d)}. Residuals of the X-ray pulse arrival times in seconds from the best fit timing model shown in (c). Neither of the two timing models provide a statistically acceptable fit to the data with a reduced $\chi^2$ of 3.2 and 2.7, respectively.}
\label{fig:timModF2F3}
\end{center}
\end{figure*}

The pulse arrival time from October 6 to November 27 can be well fit with a simple model for pulse evolution following 

\begin{equation}
\phi(t) = \phi_0 + \nu(t-t_0) + \frac{1}{2}\dot{\nu}(t-t_0)^2 + \frac{1}{6}\ddot{\nu}(t-t_0)^3 + \ldots
\label{eqTimMod}
\end{equation}

\noindent truncated at the second frequency derivative term $\ddot{\nu}$. Yet, the pulses from the early October \xmm\ observation, which starts 5 and ends 3.5 days prior to the \nicer\ October 6th observation, are offset from the simple timing model by more than half a rotation (Figure~\ref{fig:antGli}). The above simple model evidently fails to predict these pulse-phases. Hence, we added a fourth term ($\dddot{\nu}$) and fit the phase-offsets; this too does not result in a satisfactory fit showing strong residuals throughout the baseline (Figure~\ref{fig:timModF2F3}, left panels). This model results in a reduced $\chi^2$ of 3.2 for 14 degrees of freedom (dof). Simultaneously adding a fourth and a fifth ($\ddddot{\nu}$) term to equation~\ref{eqTimMod} improves the quality of the fit somewhat, resulting in a reduced $\chi^2$ of 2.7 for 13 degrees of freedom (dof), while strong phase residuals are still evident (Figure~\ref{fig:timModF2F3}, right panels).

The large and sharp October 1st and 2nd ToAs offset of more than half a cycle and the inadequacy of describing the full baseline utilizing a smooth timing model (even when instantaneously adding two extra terms to equation~\ref{eqTimMod}) is strongly suggestive of an abrupt change to the spin-frequency $\nu$, i.e., a glitch. To test this hypothesis, we added a glitch model to the smooth pulse-arrival-time evolution that described the October 6 to November 27 ToAs, whereas at $t<t_{\rm g}$

\begin{equation}
\nu(t)=\nu_{\rm t}+\Delta\nu.
\label{eqGliMod}
\end{equation}

\begin{figure*}[h!]
\begin{center}
\includegraphics[angle=0,width=0.9\textwidth]{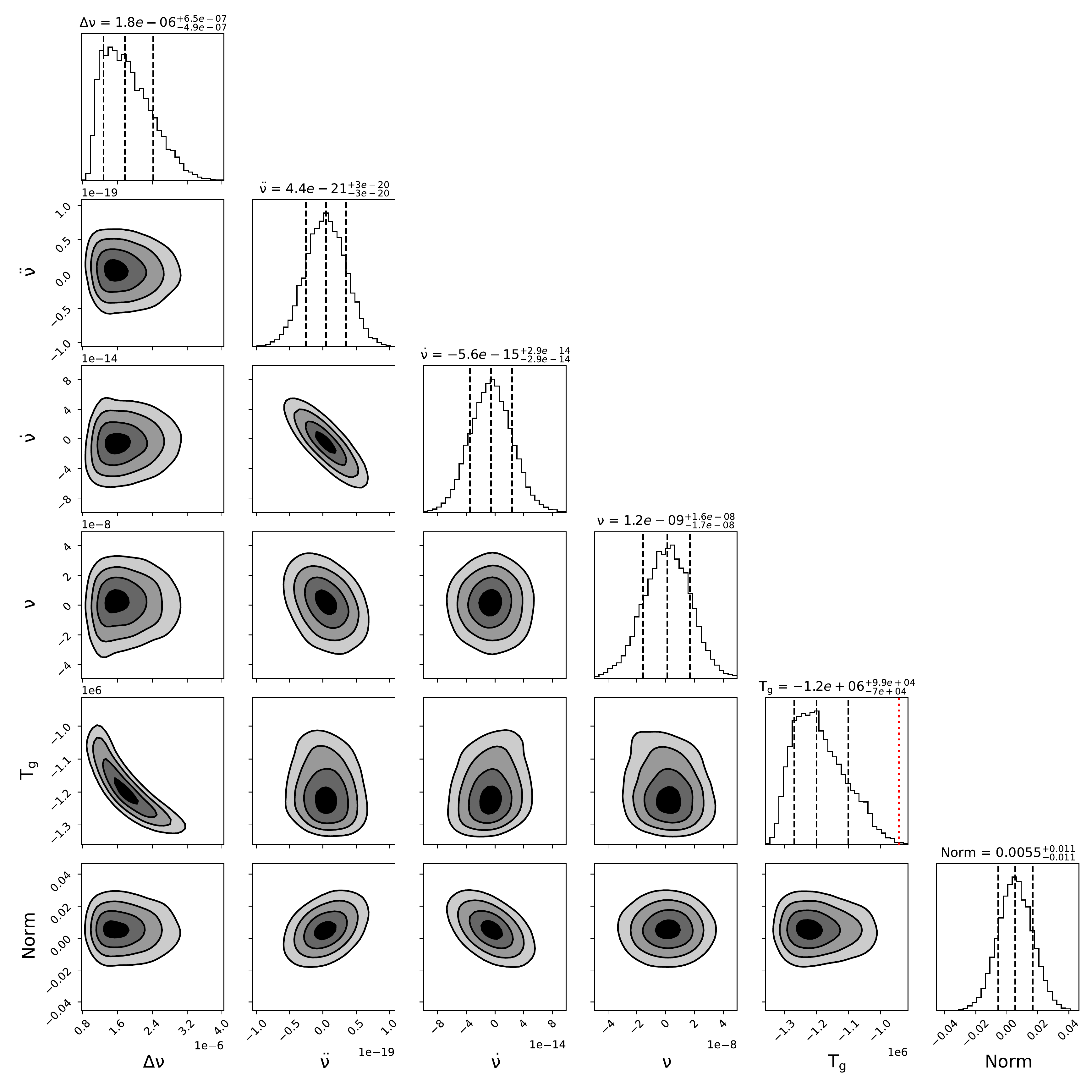}
\spacingNew{1}
\caption{One- and two-dimensional posterior probability density distributions from a 10,000 step run of the emcee sampler of the parameters of our timing model (including a glitch). The full model includes 6 free parameters; $\nu$, $\dot\nu$, $\ddot\nu$, and a normalization factor (Norm) that models the smooth evolution of the X-ray pulse arrival time, while $\Delta\nu$ and $T_{\rm g}$ are the sudden spin-frequency jump and its epoch (in seconds from 2020 October 19), respectively. The positive anti-glitch is required to predict the earlier, October 1 and 2, data with respect to the timing epoch, implying a sudden decrease of the spin-frequency at $T_{\rm g}$, i.e., an anti-glitch. Note that $\nu$, $\dot\nu$, $\ddot\nu$, and Norm are relative to the best fit model subsequent to the glitch epoch. In the 1-D histograms, the dashed lines represents the best-fit value along with its $1\sigma$ standard deviation. The dotted red line in the $T_{\rm g}$ histogram denotes the CHIME burst occurrence time.}
\label{fig:antGliCP}
\end{center}
\end{figure*}

Here $t_{\rm g}$ represents the glitch epoch, $\nu_{\rm t}$ is the predicted spin frequency subsequent to the glitch, and $\Delta\nu$ is the frequency jump at $t_{\rm g}$. The best-fit model that describes the data is shown as a dashed black line in Figure~\ref{fig:antGli}. This model results in a reduced $\chi^2$ of 1.15 for 13 dof; largely preferable to the above two timing models. Note, that this model has the same number of parameters as the continuous one with frequency derivatives up to $\ddddot{\nu}$. It is also fully consistent with the radio timing solution as derived with FAST at the $1\sigma$ level (Zhu et al. in prep., obtained through private communication). We sample the full parameter space of this model through the \texttt{emcee} MCMC sampler, assuming flat prior probability densities of all parameters and evolving 32 walkers for 10000 steps. We also exclude 500 burn-in steps. The 1- and 2-D posterior probability densities are shown in Figure~\ref{fig:antGliCP}.

\begin{figure*}[h!]
\begin{center}
\includegraphics[angle=0,width=0.98\textwidth]{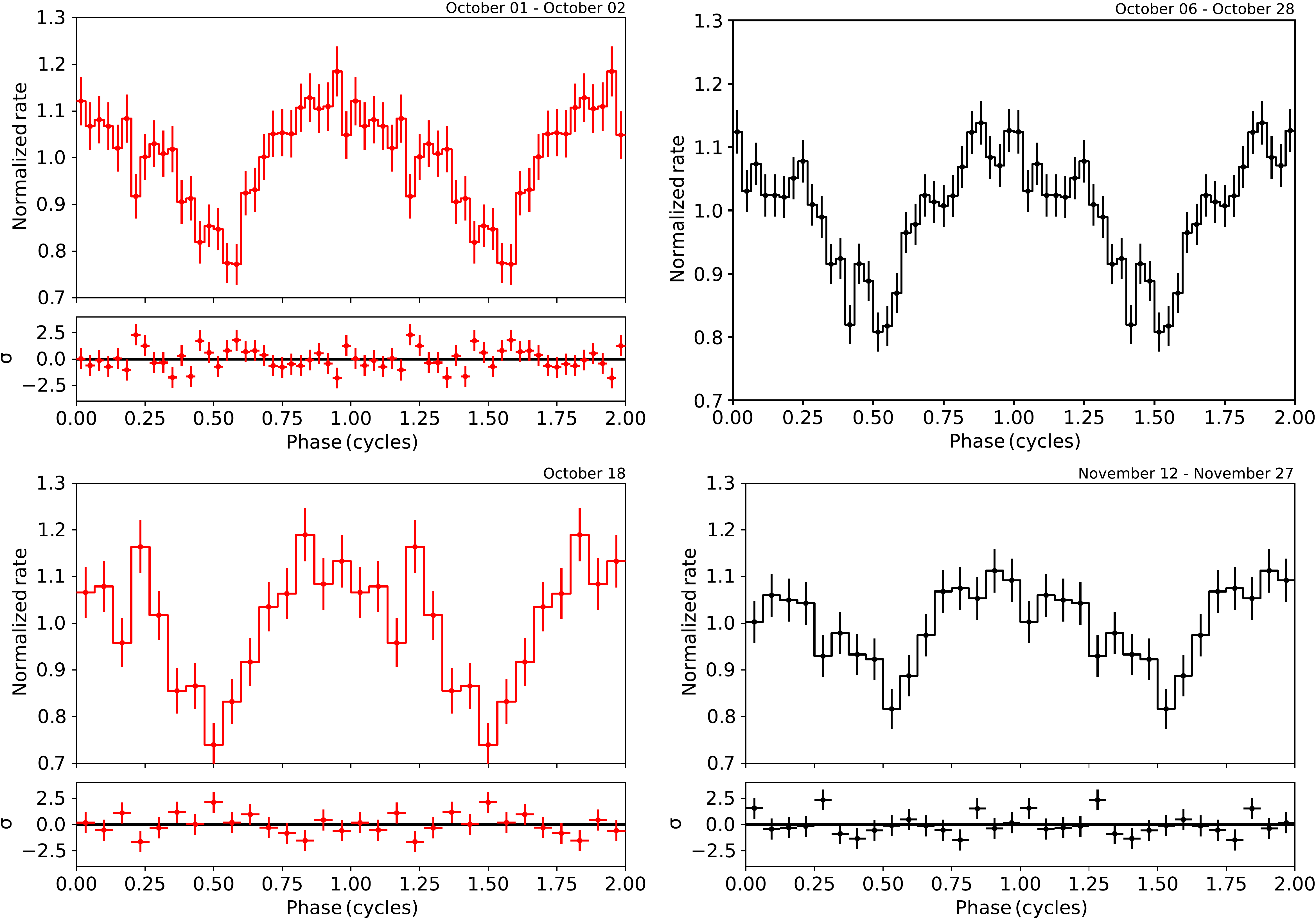}
\spacingNew{1}
\caption{Pulse profiles in the 1-3 keV energy band at different epochs during our October/November monitoring. These are folded utilizing the timing solution presented in Table~\ref{tabTimSol}. Time is indicated at the upper-right corner of each panel. Two cycles are shown for clarity. The red profiles were constructed from \xmm-only data, while the black profile are from \nicer. The source pulse profile from the \xmm\ November 12 observation is poorly constrained due to low S/N, hence, not shown. Notice the complexity of the profile, especially during the October 6 to 28 profile which boasts the largest S/N. The lower-panels of the \xmm\ and the \nicer\ November profiles are their respective deviation, in units of $\sigma$, from the high S/N October 6 to 28 \nicer\ profile, indicating no significant variability in the pulse shape throughout the validity period of our timing solution.}
\label{fig:timEvol}
\end{center}
\end{figure*}

We checked the constancy of the pulse profile shape throughout the validity period for our timing solution, most importantly to ensure that the October 1st and 2nd pulse-arrival-time deviations are not due to such variations. For this, we subtracted the October 1st/2nd \xmm\ pulse profile from the high S/N October 6 to 28 \nicer\ profile. We performed the same operation on the October 18 \xmm\ profile and the November \nicer\ profile. All the residuals are consistent with a horizontal line model ($y=$constant) having a reduced $\chi^2\approx1$. This is demonstrated in Figure~\ref{fig:timEvol} where we show the deviations, in units of $\sigma$, of these profiles from the high S/N one. Hence, we safely conclude that a change in the pulse shape cannot be attributed to the pulse-arrival-time residuals.

\begin{figure*}[h!]
\begin{center}
\includegraphics[angle=0,width=0.6\textwidth]{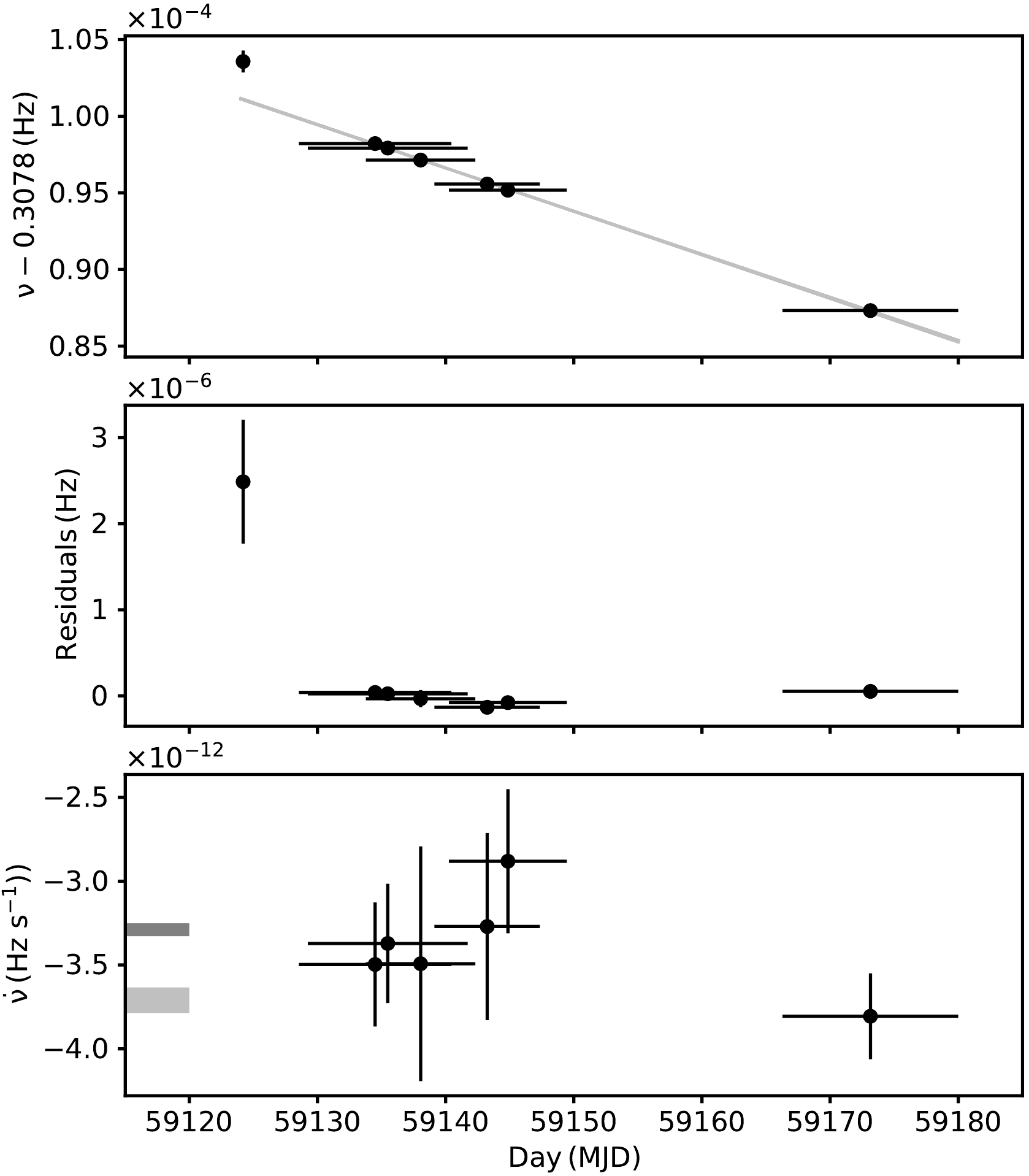}
\spacingNew{1}
\caption{Evolution of the timing properties of \src\ from 2020 October 1 to November 27. {\sl Upper-panel.} Spin frequency evolution as derived from phase-coherent timing analysis of, mostly overlapping, time segments spanning $\sim$2 weeks each. The October 1-2 spin (leftmost data point) was derived independently. {\sl Middle-panel.} Same as above, after subtracting a linear trend that best fits the 2020 October 6 to November 27 spin evolution. {\sl Lower panel.} Evolution of the spin-down rate measured from the same analysis. The light gray and dark gray bands are the spin-down rate and the corresponding uncertainty derived from phase-coherent analysis of heavy cadence \nicer\ observations covering 2020 May 19 to June 6\cite{younes20ApJ1935}, and 2020 June 18 to August 6 (Table~\ref{tabTimSol2}), respectively.}
\label{fig:testNuOct1}
\end{center}
\end{figure*}

As an extra layer of verification of our results, we performed another set of timing analysis. Over the 2020 October 6 to November 27 time period, we independently generate a phase-coherent timing solution of, mostly overlapping, time segments spanning $\sim$2 weeks each\cite{dib14ApJ}. From this time-span, we derive a set of $\nu$-$\dot{\nu}$. We also independently derive the source spin frequency during the \xmm\ October 1 data, first from a $Z^2_2$ search, then refining it through a phase-coherent analysis. These results are summarized in Figure~\ref{fig:testNuOct1}. The upper panel shows the spin evolution while the middle panel presents the residuals after subtracting a linear trend that best fits the October 6 to November 27 data (shown as a gray solid line in the upper-panel). The October 1 spin frequency is evidently incompatible with the extrapolation of the October 6-November 27 spin-evolution at the $\gtrsim3\sigma$ level, with $\Delta\nu=2.4(7)\times10^{-6}$~Hz. Note that the second largest deviation from the linear trend is $1.4\times10^{-7}$~Hz. This independent measurement of $\Delta\nu$ is consistent with the full phase-coherent analysis and, indeed, implies an abrupt slow-down on the timescale of $<$3.5 days.

Few magnetars are known to exhibit extreme timing noise especially in the months following a major outburst, e.g., 1E~1048.1$-$5937\cite{archibald20ApJ:1048} and Swift J1818.0$-$1607\cite{hu20ApJ:1818}. In these cases, the spin-down rate is erratically varying over the course of several months, reaching a maximum of around an order of magnitude larger than the nominal value. This is in contrast to the case of \src. As we show in the lower panel of Figure~\ref{fig:testNuOct1}, the spin-down rate for the two months following the spin-down glitch shows very little variability, constrained to $\lesssim20\%$. We also derive, through a phase-coherent analysis, the spin ephemerides from a heavy cadence \nicer\ observing run covering 2020 June 18 to 2020 August 6  (Table~\ref{tabTimSol2} and dark gray bar in the lower panel of Figure~\ref{fig:testNuOct1}), and show the spin-down rate measured from 2020 May 19 to 2020 June 6\cite{younes20ApJ1935} (light gray bar). The $\dot\nu$ for both epochs are consistent with the spin-down rate measured during the later 2020 period, implying that the source shows low level timing noise over longer time-scales of months.

Using the best-fit timing model for the full period, we measure the root-mean-square pulsed fraction in the energy range 1-3~keV of several combined \nicer\ data sets to boost the S/N and individual \xmm\ observations using a Fourier series consisting of two harmonics\cite{woods04ApJ:1E2259}. We find a stable pulsed fraction of around $(9\pm1)\%$ (panel (c) in Figure~\ref{fig:antGli}). We find no strong pulsed emission at energies $>3$~keV in either instrument, nor did we find any pulsed emission in \nustar. For the October 1st data which boasts the highest S/N at energies $>3$~keV, we derive a $3\sigma$ upper-limit of $12\%$ on the pulsed fraction at energies 3-10 keV. We also built a \nicer\ pulse profile for the October 6 to 28 period which overlaps with the radio-pulsar period (Figure~\ref{fig:timEvol}). We note the complexity of the profile. The multitude of peaks and their spread across rotational phase are highly indicative of large portions of the star surface being activated.

\begin{figure*}[h!]
\begin{center}
\includegraphics[angle=0,width=0.48\textwidth]{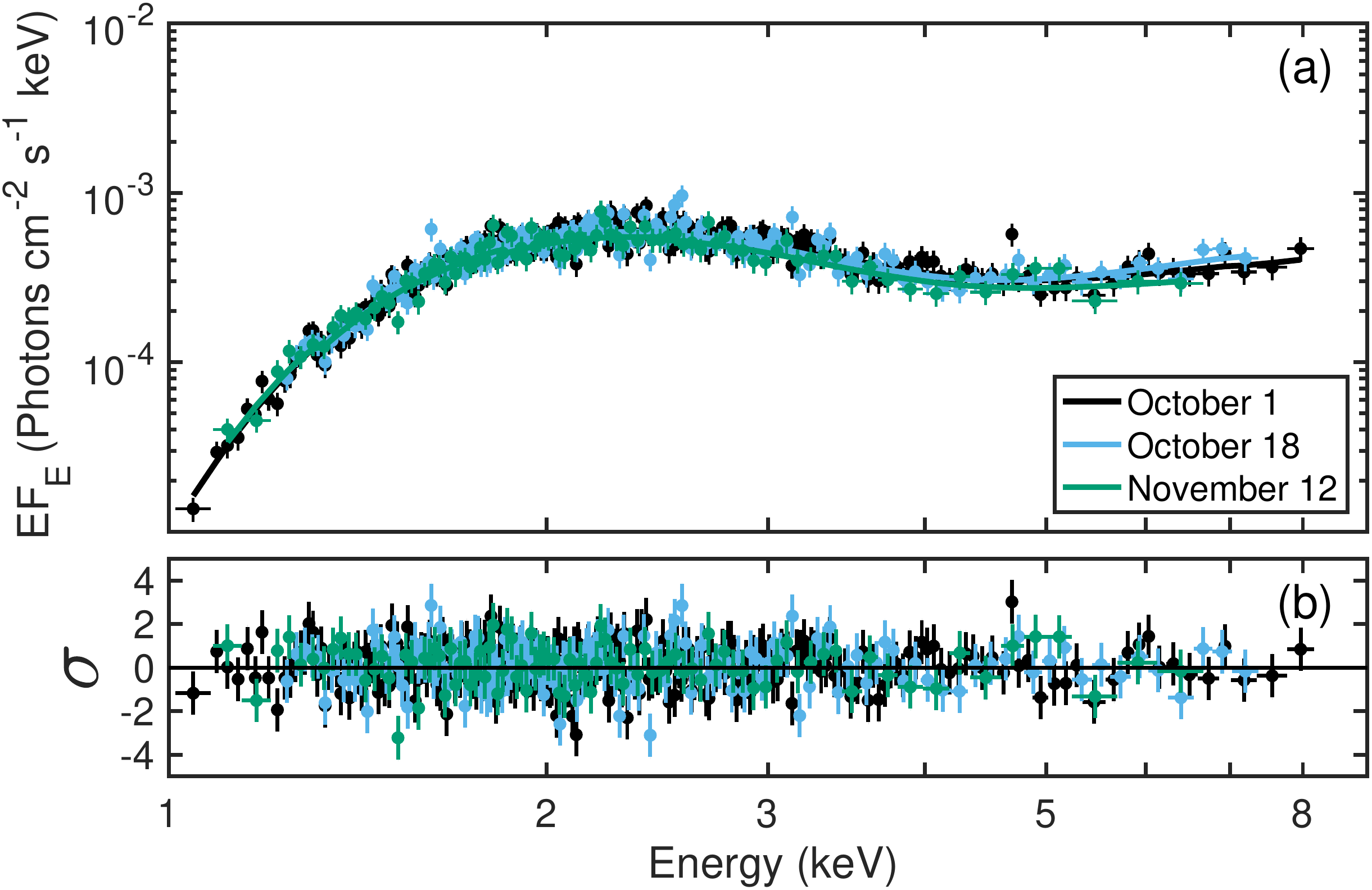}
\includegraphics[angle=0,width=0.48\textwidth]{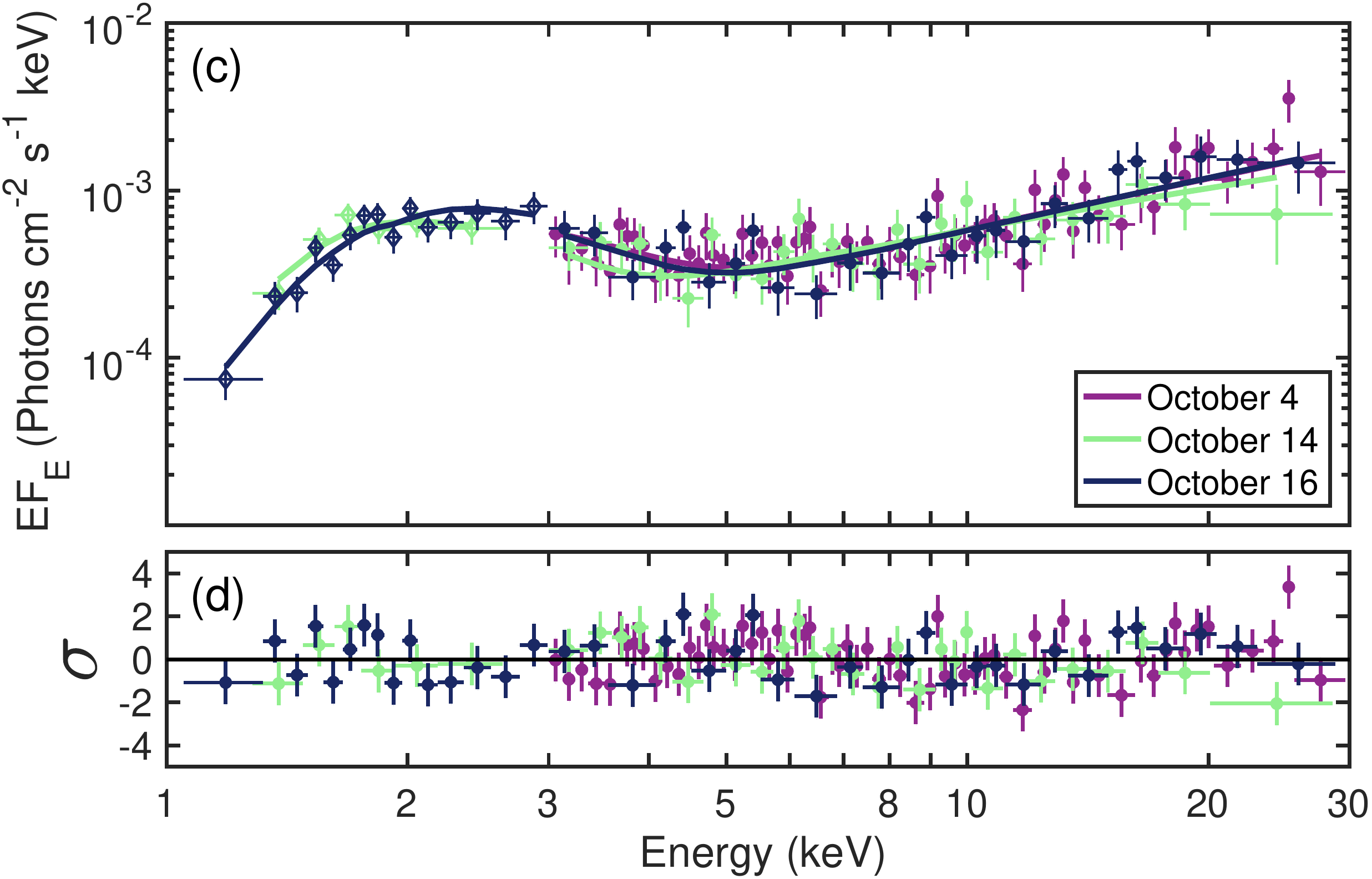}
\spacingNew{1}
\caption{{\sl Panel (a)}. X-ray spectra of the \xmm\ observations. The dots represent the data while the solid lines are the best-fit absorbed blackbody+power-law models. The spectra are color-coded by start-date of each observation. {\sl Panel (b)}. Residuals in units of 1 standard deviation from the best fit model. {\sl Panel (c)}. Same as (a) but for the \nustar\ and (quasi-)simultaneous \nicer\ observations. The colors represent the start-date of each \nustar\ observation. {\sl Panel (d)}. Residuals in units of 1 standard deviation from the best fit model. No statistically significant variability in the spectral shape is detected.}
\label{fig:specXMM}
\end{center}
\end{figure*}

\noindent {\bf Spectral analysis.} Utilizing the \texttt{grppha} command within HEASOFT, we group the \xmm\ and \nustar\ spectra to have at least 30 counts per energy bin, and 5 counts per energy bin for \nicer. We fit the three \xmm\ spectra in the 1-9 keV range simultaneously with a combination of a thermal blackbody and power-law components, both affected by interstellar absorption. We leave all model parameters free to vary, except for the hydrogen column density $N_{\rm H}$ of the interstellar absorption model. As shown in Figure~\ref{fig:specXMM}, the model describes the data well with no clear systematic residuals. The reduced $\chi^2$ is approximately $1.0$ for 730 degrees-of-freedom (dof). The best-fit model parameters along with their uncertainties are summarized in Table~\ref{specParam}. There is no significant variation in any of the model parameters between the three epochs, albeit sampling the full October and November time period, which coincide with the pre-glitch, post-glitch and radio-turn on, and the radio turn-off of November. This demonstrates the stability of the soft and hard X-ray emitting region throughout the glitch/radio-on time interval.

We fit the \nustar\ spectra of the three different observations to the same model as above in the 3~keV to 30~keV energy range. We supplement the \nustar\ spectrum of October 16 with a simultaneous \nicer\ observation (observation ID 3020560159), and the one from October 14 with a quasi-simultaneous \nicer\ observation (observation ID 3020560158) taken 22 hours before. We link the hydrogen column density amongst all spectra while allowing the rest of the model parameters to vary. The model adequately fits the data with a reduced $\chi^2$ of 0.9 for 659 dof. We summarize the spectral results in Table~\ref{specParam}. The \nustar+\nicer\ spectra confirm the stability of the thermal and non-thermal components. The October 4th \nustar\ observation, which ended at the best-fit glitch epoch, places stringent constraints on the time-scale of any induced radiative variability.

Finally, we fit the individual \nicer\ observations to the same model as above in the 1~keV to 5~keV energy range. The \nicer\ spectra cannot constrain the power-law component due to the low S/N at energies $>3$~keV. Hence, we fixed the power-law model parameters to those of \xmm. Moreover, we linked the blackbody temperature between all spectra since it shows no sign of significant variability. We present the 1-10 keV flux evolution throughout the full October and November period in Figure~\ref{fig:antGli}.

\noindent {\bf Burst searches.} We employ a Poisson methodology to search for any bursts in all of our data-sets\cite{gavriil04ApJ:1E2259, younes20ApJ1935}. In summary, using a 32~ms binned light curve, we flag any deviation from the average count rate in a single observation that cannot be ascribed to random Poisson fluctuation. We then scrutinize these bins to eliminate spurious detections, e.g., due to flaring background. We do not find any bursts with significance $>5\sigma$ in any of the observation. We repeated our procedure for different time-scales, namely 64, 128, 512, and 1024~ms and found no burst-candidates. The average \nicer\ count rate of \src\ in the 1-8 keV is $\approx0.7$~counts~s$^{-1}$. Assuming a top-hat burst with duration of 1~s, we place a $5\sigma$ detection upper-limit of about 8~counts~s$^{-1}$, which translates to an absorption corrected flux of $10^{-10}$~erg~s~$^{-1}$~cm$^{-2}$ assuming $N_{\rm H}=2.2\times10^{22}$~cm$^{-2}$ and a blackbody spectrum having a temperature  $kT=1.5$~keV\cite{younes20ApJ1935}. At a distance of 10~kpc\cite{zhong20ApJ1935}, this implies a burst luminosity of $10^{36}$~erg~s$^{-1}$.

\noindent {\bf Chance alignment of an anti-glitch, radio bursts, and pulsed-radio episode.} Due to the lack of years-long monitoring of \src\ with either X-ray or radio facilities, and our poor knowledge of the waiting-time distribution of any of these phenomena, we do not attempt to provide a quantitative measurement of the probability of the three events occurring within few days of one another. Rather, we provide a qualitative description of their occurrence rate and argue of their individual rarity. 

\src\ has been coherently timed on two occasions, 2014 July-2014 November\cite{Israel-2016-MNRAS} and our current epoch covering 2020 October-November, for a total of 180 days. Any spin-down glitch with a magnitude similar to the one presented in Table~\ref{tabTimSol} would be easily detectable during the above two-periods, implying a rough upper limit of one spin-down glitch every 0.5 year. We also note that such events are exceedingly rare within the magnetar population, having been conclusively detected in only one other magnetar 1E 2259+586. In the latter, three such events have been detected in the course of 20 years of monitoring\cite{dib14ApJ,younes20ApJ1935} implying a rate of at most one in 6 years. 

The CHIME radio dishes have good daily coverage of \src\ totalling 15 minutes. Including the detection of the FRB on April 28, this translates to a rate of roughly $4\times10^{-4}$ radio bursts per CHIME-day in 2020 (less if one considers CHIME observations in 2018 and 2019, yet \src\ was mainly quiet during these years). Again, note that this assumes a Poisson process. From continuous radio observations of \src\ in the weeks and months following the April 28 event, it is evident that these radio bursts are rare and highly clustered\cite{lin2020Natur,kirsten2020,bailes21mnras}. With this caveat in mind, we derive a joint probability of about $10^{-5}$ that the anti-glitch and the CHIME bursts occur within a 3-day period, corresponding to a $4.2\sigma$ chance coincidence.

The faintness and narrow frequency range of the \src\ radio pulsed emission makes it difficult to detect with radio dishes apart from FAST (Zhu et al. in prep., obtained through private communication). Nevertheless, FAST performed regular observations of \src\ between 2020 April 15 and 2021 July 14 at a cadence varying from once per day to about once per month. The only period with a pulsed radio detection is the October 9 to 28. The earliest FAST observation to the radio turn-on occurred on 2020 August 28, i.e., 35-days prior to our spin-down glitch epoch. 

The transient nature of the pulsed radio emission in magnetars argues that it must be triggered by a certain event, which likely causes magnetospheric modifications connected to the open field line region. The \src\ long term light curve does not show any notable change to the X-ray spectral properties in 2020 August and September\cite{younes20ApJ1935,borghese22MNRAS}, nor were there any reported hard X-ray bursts from large field of view telescopes, e.g., Fermi/GBM or Swift/BAT. The only notable anomaly in this time-period is the spin-down glitch. Hence, while it is impossible to prove that no radio pulsations occurred during the radio-dark 35-day interval, the fact that the spin-down glitch is the only notable spectral or temporal anomaly surrounding the observed radio activation is strongly suggestive of an association of these two exceedingly rare events, rather than them being a chance coincidences.

\noindent {\bf Ephemeral wind interpretation.} We explore here how mass loss through a transient, strong wind that extracts angular momentum from the star is constrained through the spin-down glitch.
Such an external process is perhaps a most natural interpretation in considering physical connections of an abrupt spin-down to an incipient radio signal.  A sudden (small) increase in stellar oblateness in dimensions perpendicular to the spin axis 
could explain the anti-glitch, which could be effected by a changing magnetic ``buoyancy'' in the crust \cite{Mastrano-2015-MNRAS}. Standalone, it is unclear how this would precipitate a magnetospheric radio signal.  Yet it could arise in conjunction with a magnetic energy release near the poles that drives a wind that we now describe.

\begin{figure*}[h!]
\begin{center}
\includegraphics[angle=0,width=0.95\textwidth]{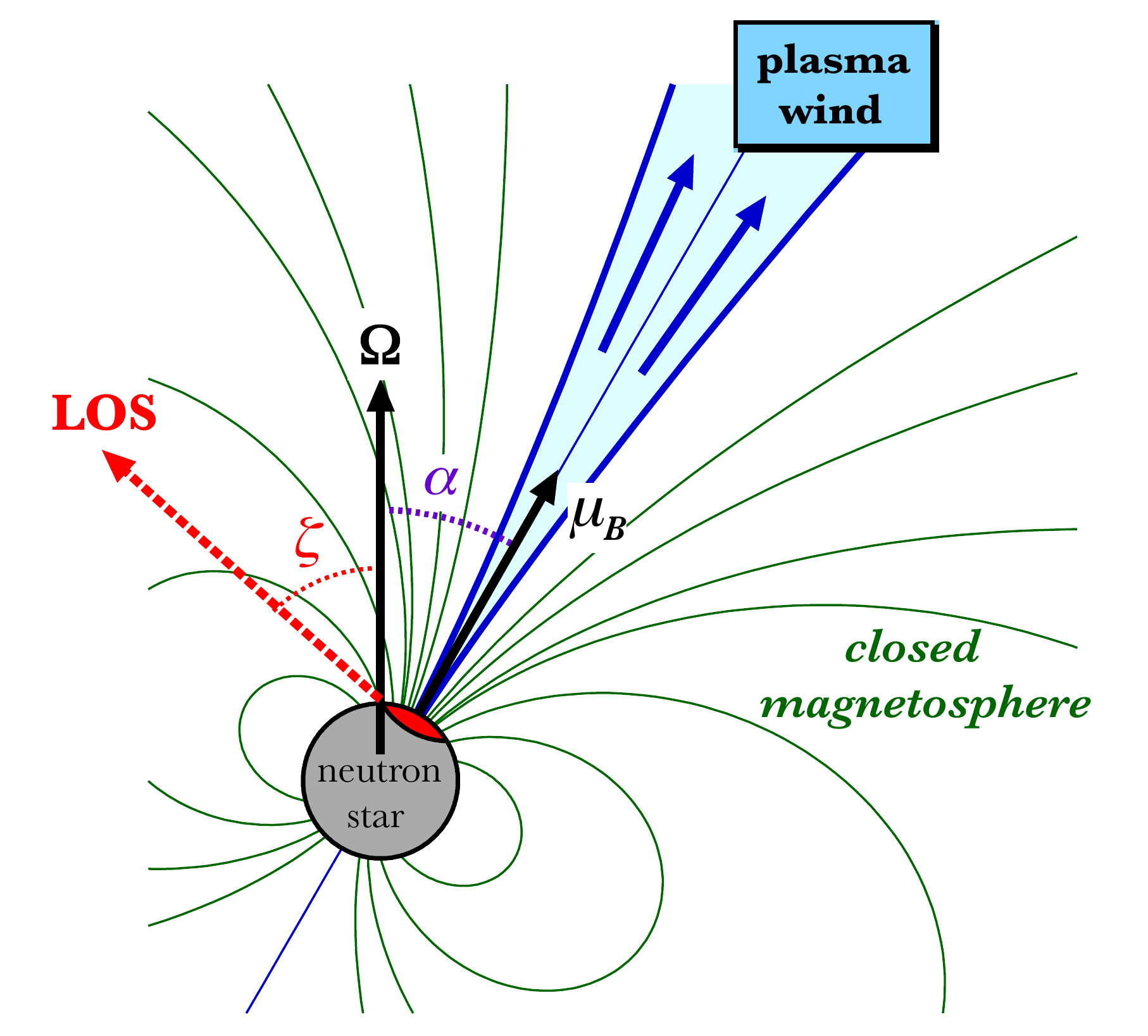}
\spacingNew{1}
\caption{Schematic diagram showing the stellar geometry during the ephemeral wind epoch. Two angles are marked: $\alpha$ between the neutron star's magnetic ($\boldsymbol{\mu}_B$) and rotation ($\boldsymbol{\Omega}$) axes, and $\zeta$ between $\boldsymbol{\Omega}$ and the observer's line of sight (LOS) to the hot polar cap (red) of persistent soft X-ray emission (employed in the timing solutions); these angles are fixed on long timescales. The polar region is shaded in light blue; it contains highly-twisted, quasi-radial, open field lines depicted in dark blue. The remaining magnetosphere is closed, with field lines in dark green that are quasi-dipolar near the magnetic equator, yet can possess moderate twists near the periphery of the wind zone. The bulk of the ephemeral wind flows along the open field line region, though a small portion could flow along quasi-polar closed field lines.}
\label{fig:wind_geom}
\end{center}
\end{figure*}

The basic geometry of the wind and stellar configuration is depicted in Figure~\ref{fig:wind_geom}.
Let \teq{\delta m \sim {\dot m}\delta t} be the cumulative mass shed in time \teq{\delta t}, putatively at an approximately 
constant rate \teq{\dot m}, on field lines with footpoints
very near the magnetic pole.  If the 
inclination angle between the magnetic and rotation axes is \teq{\alpha}, then 
\teq{R\Omega\sin\alpha} is the circular rotation speed 
at altitude \teq{R} above the magnetic pole.  Therefore the angular momentum shed is of the order of 
\teq{\delta m R^2 \Omega\sin\alpha}.  As the wind flows out towards the light cylinder, the 
star continues to transfer angular momentum to the wind until the magnetic energy density 
drops below that of the plasma at \teq{R\sim R_{\rm eq}}. 
Subsequently, the wind combs the field lines out and the wind's angular 
momentum decouples from the stellar rotation.   The net angular momentum transfer 
from the magnetar to the wind is 
\teq{\delta {\cal L}_{\rm w} = \delta m R_{\rm eq}^2 \Omega\sin\alpha}: it
can be equated to \teq{I \vert \delta\Omega\vert}, where \teq{\delta\Omega} is the 
abrupt change in the rotation frequency measured by the timing data.
Assume that  the star's moment of inertia 
\teq{I = 2\epsilon MR_{\rm ns}^2/5} is essentially constant during the shedding event,
where \teq{\epsilon} represents the departure from a uniform density sphere.  Thus, 
\begin{equation}
   \frac{\delta m}{M_{\rm ns}} \; \sim\; \dover{2\epsilon}{5} \, \frac{R_{\rm ns}^2}{R_{\rm eq}^2} \, 
   \frac{\vert\delta \Omega \vert}{\Omega\sin\alpha } \quad .
 \label{eq:mass_shed_method}
\end{equation}
The spin-down glitch establishes \teq{-\delta \Omega/\Omega = 5.8 \times 10^{-6}}.

To determine the equipartition radius \teq{R_{\rm eq}} at which the plasma energy 
density begins to exceed the magnetic field energy density \teq{B^2/8\pi}, 
presume that the wind flows with a mean wind speed of \teq{\beta_{\rm w}c}, 
with \teq{\beta_{\rm w}\lesssim 0.8} that is mildly-relativistic.  At altitude \teq{R_{\rm eq}}, the cross sectional 
area of the open field line flux tube is \teq{A = \pi R_{\rm eq}^2\theta^2}, with 
\teq{R_{\rm eq}/\theta^2 \sim R_{\rm lc} = Pc/2\pi} defining the local colatitude \teq{\theta} (\teq{\ll 1})
of the last open field line in a dipolar geometry.  Note that plasma loading of the 
magnetosphere will enlarge this area of open field lines\cite{Harding-1999-ApJ}.  
The mass flux through this area couples to the mass 
density \teq{\rho} via the conservation relation \teq{{\dot m} = \rho  A \beta_{\rm w}c}.
Equating \teq{\rho c^2} to \teq{B^2/8\pi} at altitude 
\teq{R_{\rm eq}} gives \teq{\rho\to \rho_{\rm eq}} and
\begin{equation}
   \rho_{\rm eq} c^2 \; \sim\; \dover{{\dot m} c^2P}{2 \pi^2 R_{\rm eq}^3\beta_{\rm w}}
   \; \sim\; \dover{B_p^2}{8\pi} \left( \dover{R_{\rm ns}}{R_{\rm eq}} \right)^6 \quad .
 \label{eq:equipartition_method}
\end{equation}
Here, \teq{B_p} is the surface polar field strength.  This constrains \teq{R_{\rm eq}}, 
and when combined with the angular momentum budget equation 
in Eq.~(\ref{eq:mass_shed_method}) to eliminate \teq{\delta m} yields
\begin{equation}
   \dover{R_{\rm eq}}{R_{\rm ns}} \;\sim\; 
   \left( \dover{\pi}{4} \, \dover{B_p^2 R_{\rm ns}^3 \beta_{\rm w}}{\delta m\, c^2} 
      \, \dover{\delta t}{P} \right)^{1/3} 
   \;\sim\; \lambda \, \dover{\Omega}{\vert \delta\Omega\vert} 
           \, \dover{B_p^2 R_{\rm ns}^3}{M c^2} \, \dover{\delta t}{P} \quad ,
 \label{eq:Req_est_final}
\end{equation}
with \teq{\lambda = (5\pi / 8 \epsilon) \, \beta_{\rm w}\sin\alpha}.
Inserting this into Eq.~(\ref{eq:mass_shed_method}) delivers the fractional
stellar mass \teq{\delta m/M} lost to the ephemeral wind.  Evaluation for 
a transient wind duration of 10 hours yields
\begin{equation}
   \dover{\delta t}{P} \; =\; 1.1 \times 10^4 \; \rightarrow \; \dover{R_{\rm eq}}{R_{\rm ns}} \;\sim\; 150
   \quad \hbox{and}\quad 
   \dover{\delta m}{M} \; \sim\; 10^{-10} \quad .
 \label{eq:RSDE_soln_method}
\end{equation}
This estimate assumes that \teq{\lambda \sim 1}, and that \teq{\epsilon/ \sin\alpha =1}.
The result is a modest fractional mass loss (i.e., 
\teq{\sim 2.5 \times 10^{44}}~erg in total energy with a wind luminosity of 
\teq{L_{\rm w}\equiv {\dot m}c^2 \sim 7\times 10^{39}}~erg~s$^{-1}$) that implies that spin-down glitches can be 
recurrent events on timescales of 10 years or more for magnetars throughout 
a putative \teq{10^4}year lifetime.  Their polar confines suggest a precipitating event somewhat akin to volcanism, 
spewing out plasma at mildly-relativistic speeds.
This could be driven via disruptive magnetic stress and energy release in the crust.
To contrast, rapid mass loss on a timescale of a few minutes (\teq{\delta t/P = 10^2})
implies \teq{R_{\rm eq}\sim 1.5 R_{\rm ns}} and \teq{\delta m/M \sim 10^{-6}}
for \teq{M=1.4 M_{\odot}}.  
This circumstance essentially approximates a structural rupturing of the outer crust,
with a wind luminosity of \teq{L_{\rm w} \sim 7\times 10^{45}}erg/sec
that is comparable to the radiative luminosity of the ``initial spike''
of a magnetar giant flare.

The above calculations constitute an approximate upper bound to the mass loss
and wind luminosity for fixed \teq{\delta t}.  The dipole configuration used therein is an idealized 
choice that needs to be adapted to treat more realistic descriptions of plasma-loaded 
magnetospheres such as in pulsar plasma simulations\cite{Bai-2010-ApJ,Kalapotharakos-2012-ApJ} and magnetar analytic 
models\cite{Harding-1999-ApJ}.  The upshot of plasma loading is that it expands the zone of open field 
lines,  so that the dipole form \teq{\theta \sim [R_{\rm eq}/R_{\rm lc}]^{1/2}} is an 
underestimate for \teq{\theta}, and reduces the size of the magnetosphere.
Without introducing extra parametric complexity, the quickest way to get a sense of this 
plasma-driven opening of the magnetosphere is to note that it is akin to shortening 
the rotation period \teq{P}.  Thus, for example, lowering \teq{P} in 
Eq.~(\ref{eq:Req_est_final}) by a factor of \teq{10}, increases \teq{R_{\rm eq}}
by the same factor, and then reduces the net mass loss in Eq.~(\ref{eq:mass_shed_method})
by two orders of magnitude, and accordingly would result in 
\teq{L_{\rm w}\sim 7\times 10^{37}}erg/sec for \teq{\delta t=10}hours.  This 
drop is driven by the lower densities in Eq.~(\ref{eq:equipartition_method})
required to realize a given angular momentum shed at high altitudes.  Accordingly, 
plasma loading influences on the field structure, and likewise twist modifications, 
will generally lower the average mass loss rates.

\noindent{\bf Opacity of the wind.}
Such a dense wind has the potential to occult the surface and magnetospheric signals.  
Yet, the \nustar\ and \nicer\ observations indicate that such an obscuration is not significant.
One can quickly estimate the lepton number density \teq{n_e = \rho /m_e} in a pure pair plasma in the wind zone out to 
any radius \teq{R\lesssim R_{\rm eq}\ll R_{\rm lc}} along the open field lines.  
For the plasma, the flared wind solution \teq{A = \pi R^2\theta^2}, with 
\teq{R/\theta^2 \sim R_{\rm lc} = Pc/2\pi \approx 1.5\times 10^{10}}cm, yields 
a radial dependence of \teq{\rho = \rho_{\rm eq} \, (R_{\rm eq}/R)^3}. 
The non-magnetic Thomson opacity \teq{\taut = n_e \sigt R}, appropriate for 
the sub-critical fields at \teq{R\gtrsim 10R_{\rm ns}}, can be developed 
using Eq.~(\ref{eq:equipartition_method}), yielding
\begin{equation}
   \taut \; =\; \sigt R_{\rm ns} \, \dover{B_p^2}{8\pi m_ec^2} 
   \left( \dover{R_{\rm ns}}{R_{\rm eq}} \right)^5  \left( \dover{R_{\rm eq}}{R} \right)^2 
   \quad , \quad
   R \;\lesssim\; R_{\rm eq} \quad .
 \label{eq:tau_Thomson_method}
\end{equation}
For the wind configuration given by Eq.~(\ref{eq:RSDE_soln_method}), 
this pair plasma result evaluates to \teq{\sim 8.2\times 10^4} 
at \teq{R_{\rm eq}}, i.e., an extremely high opacity, and the wind remains 
opaque out to beyond the light cylinder.  If instead, the plasma 
is hydrogenic, the optical depth is reduced by a factor of \teq{m_e/m_p}, 
yielding \teq{\taut \sim 50} at \teq{R_{\rm eq}\sim 150 R_{\rm ns}}. For either hydrogenic or pair plasma, \teq{\taut} is extremely large at the stellar surface.

The ephemeral wind will clearly obscure any background radiation field from the surface or inner magnetosphere that impinges upon it.  Yet the solid angle of the wind at \teq{R_{\rm eq}} is small.  Its effective area at this radius is \teq{A = \pi R_{\rm eq}^2\theta^2 \sim \pi R_{\rm eq}^3/R_{\rm lc}}, constituting a solid angle of \teq{\sim \pi R_{\rm eq}/R_{\rm lc}\sim 3\times 10^{-2}} steradians, corresponding to \teq{\theta \sim 5.7^{\circ}} for the dipole, and larger for field geometry modifications due to plasma loading. Above \teq{R_{\rm eq}}, the wind combs out the field and propagates radially, so its solid angle is approximately preserved out to the light cylinder.  Thus, wind occultation of the persistent emission from low altitudes or the surface is relatively small, even though the wind will remain optically thick out to \teq{R_{\rm lc}}. In the putative subsequent residual wind phase, when observations are resumed, \teq{\vert \delta \Omega\vert} is much smaller than during the anti-glitch. The value of \teq{\dot m} is likely at least 3-4 orders of magnitude smaller, \teq{R_{\rm eq}} increases by a factor of 10-20, and it can be quickly shown that while the solid angle of the residual wind is large near the light cylinder, this more benign wind is transparent to Thomson scattering there.

\noindent {\bf Conditions for pair creation}:
The historical paradigm that an abundance of electron-positron pairs is 
required for persistent radio emission in pulsars\cite{Sturrock-1971-ApJ} still 
prevails.  Radio pulse profile and polarization constraints indicate 
that the altitude of radio emission\cite{Blaskiewicz-1991-ApJ,Dyks-2004-ApJ} is generally in the 
100 - 1000 km range, and is presumed to occur over the magnetic poles.
During the strong ephemeral wind epoch, the opacity is so enormous 
that it precludes the formation of electric potential ``gaps.''   These potentials 
seed primary electron acceleration and subsequent 
curvature radiation that lead to QED magnetic pair creation \teq{\gamma\to e^+e^-}
and ultimately cascading.\cite{Daugherty1982}
After the strong wind phase has ceased, the opacity drops precipitously 
and electric potentials can stably exist, so that pair creation and radio emission may 
become possible.

As \teq{\gamma\to e^+e^-} has a fundamental energy threshold of 
\teq{2m_ec^2\sin\theta_{\rm kB}}, where \teq{\theta_{\rm kB}} is 
the angle of gamma-ray propagation relative to the local field direction, magnetic photon splitting \teq{\gamma\to\gamma\gamma}
can be a prolific competitor in magnetars\cite{Baring-1998-ApJL,Baring-2001-ApJ} since it 
has no such threshold.  In a first examination of this possibility, Baring \& Harding\cite{Baring-1998-ApJL} 
concluded that suppression of pair creation by photon splitting
is efficient in magnetars and could explain why no radio magnetars 
had been detected prior to 2000, and only a handful of transient ones since,
SGR~1935+2154 being the latest.  Yet the balance in the competition 
between pair conversion and splitting of gamma-rays depends on 
the inner magnetospheric emission locale in magnetars, with splitting tending 
to dominate in polar regions where a dipolar field is stronger, and pair creation 
being favored in non-polar locales where the field line radius of curvature 
is smaller.\cite{Hu-2019-MNRAS} 

During the strong polar wind phase, the magnetic field lines are combed out 
radially above \teq{R_{\rm eq}\gtrsim 15R_{\rm ns}}, 
similar to that evinced in pulsar magnetosphere simulations.\cite{Kalapotharakos-2012-ApJ,Tchekhovskoy-2016-MNRAS}
Vestiges of this field structure will persist for some time 
after the anti-glitch.  On the long term, magnetars are believed to possess globally twisted fields 
with toroidal components generated by surface and magnetospheric currents.\cite{Thompson-2002-ApJ} 
In the domain of very strong twists, the 
field morphology approaches a split monopole and therefore resembles plasma-loaded magnetospheric geometry.  The introduction of strong twists
moves the zones of dominance 
by pair creation by gamma-rays of energies \teq{\gtrsim 50}MeV 
more towards the poles and to slightly higher altitudes\cite{Hu-2022-ApJ}
\teq{\sim 10-30R_{\rm ns}}, which are still below the putative locales for radio emission.  This change is precipitated by the extreme sensitivity of the pair creation rate to the field strength \teq{\vert \mathbf{B}\vert} and curvature of magnetic field lines, and the direction of gamma-rays relative to {\bf B}.\cite{Harding-1978-ApJ,Baring-1998-ApJL,Baring-2001-ApJ}
Thus, we anticipate that polar pair creation and pulsed radio emission can proceed after 
the strong wind abates significantly.  Yet Ohmic dissipation of toroidal/twisted fields 
in magnetars is nominally on the timescale of months to years,\cite{Beloborodov-2009-ApJ}
depending on the voltage along the pertinent field lines.
Presuming that a similar relaxation transpires in the decaying 
wind scenario here, photon splitting will eventually again dominate 
in polar colatitudes\cite{Hu-2019-MNRAS} after sufficient untwisting, and magnetic pair creation and radio emission 
there will cease.  The magnetic evolution must be largely confined to the polar regions so as to not influence the persistent surface and hard X-ray signals substantially.

Thus the picture we envisage is that the anti-glitch
creates plasma loading with a stronger twist in polar zones that allows some transient
pair production and radio emission once the wind density drops and the outflow becomes optically thin.  Then later in the wind abatement phase, the field twist relaxes back
to its persistent twist configuration.
This conjecture motivates deeper study of field morphology and its evolution 
via Ohmic dissipation in concert with gamma-ray opacity  and pair 
creation considerations.  On the observational side, NASA's new
Imaging X-ray Polarimetry Explorer \cite{IXPE-2021-arXiv} (IXPE) could help constrain field structure in bright magnetars around (spin-down) glitch epochs through its polarization measurements of soft X-ray emission emanating from their surfaces.

\noindent {\bf Radiation from the strong wind}:
A long-lived non-thermal-like radiation signal associated with the hot wind would be expected likely spanning X rays/EUV down to the optical/IR/mm band as the wind adiabatically cools on its path out to \teq{R_{\rm lc}}. Its bolometric luminosity would be a small fraction of the wind luminosity, i.e., \teq{L_{\rm w} \lesssim 7\times 10^{39}}~erg~s$^{-1}$ for \teq{\delta t=10}~hours under the simple assumption of a pure dipole (or \teq{L_{\rm w} \lesssim 7\times 10^{37}}~erg~s$^{-1}$ for a polar cap size an order of magnitude larger, see above). The intrinsic radiation efficiency is small in wind/jet systems, of the order of 0.1-1\%, particularly if it is baryon loaded (i.e. hydrogenic), as exemplified in supernovae and gamma-ray bursts: heat plus radiation pressure is efficiently converted into bulk kinetic energy of the plasma. The high opacity of the wind just above the stellar surface will drive it towards thermal equilibrium in all but a thin outer sheath.  Using  \teq{L_{\gamma} = 4 \pi \sigma T^4 R^2\lesssim 10^{38}}~erg~s$^{-1}$, one quickly estimates the plasma/radiation temperature at the surface to be \teq{T\lesssim 2\times 10^7}K \teq{=1.6}~keV for the \teq{\delta t=10}~hour case, using the Stefan-Boltzmann law. As the wind flows to higher altitudes, it quickly cools according to the adiabatic expansion law \teq{V^{\gamma -1}T=} const., with \teq{\gamma \approx 5/3} as the ratio of specific heats. For dipole field morphology, the comoving volume of the wind is \teq{V\propto 1/\rho\propto R^3}, whereas for an isotropic wind above \teq{R_{\rm eq}}, \teq{V\propto 1/\rho\propto R^2}, respectively  yielding \teq{T\propto R^{-2}} and \teq{T\propto R^{-4/3}}. Thus, the wind temperature drops below \teq{10^3}~K by the time it reaches  \teq{R_{\rm eq}\sim 150 R_{\rm ns}}, and the ``isotropic'' optical luminosity (\teq{\propto T^4 R^2}) is \teq{10^{25}}~erg~s$^{-1}$ or much less. The small solid angle lowers the potential visibility of any radiation signal associated with the ephemeral wind. 

\begin{figure*}[h!]
\begin{center}
\includegraphics[angle=0,width=0.95\textwidth]{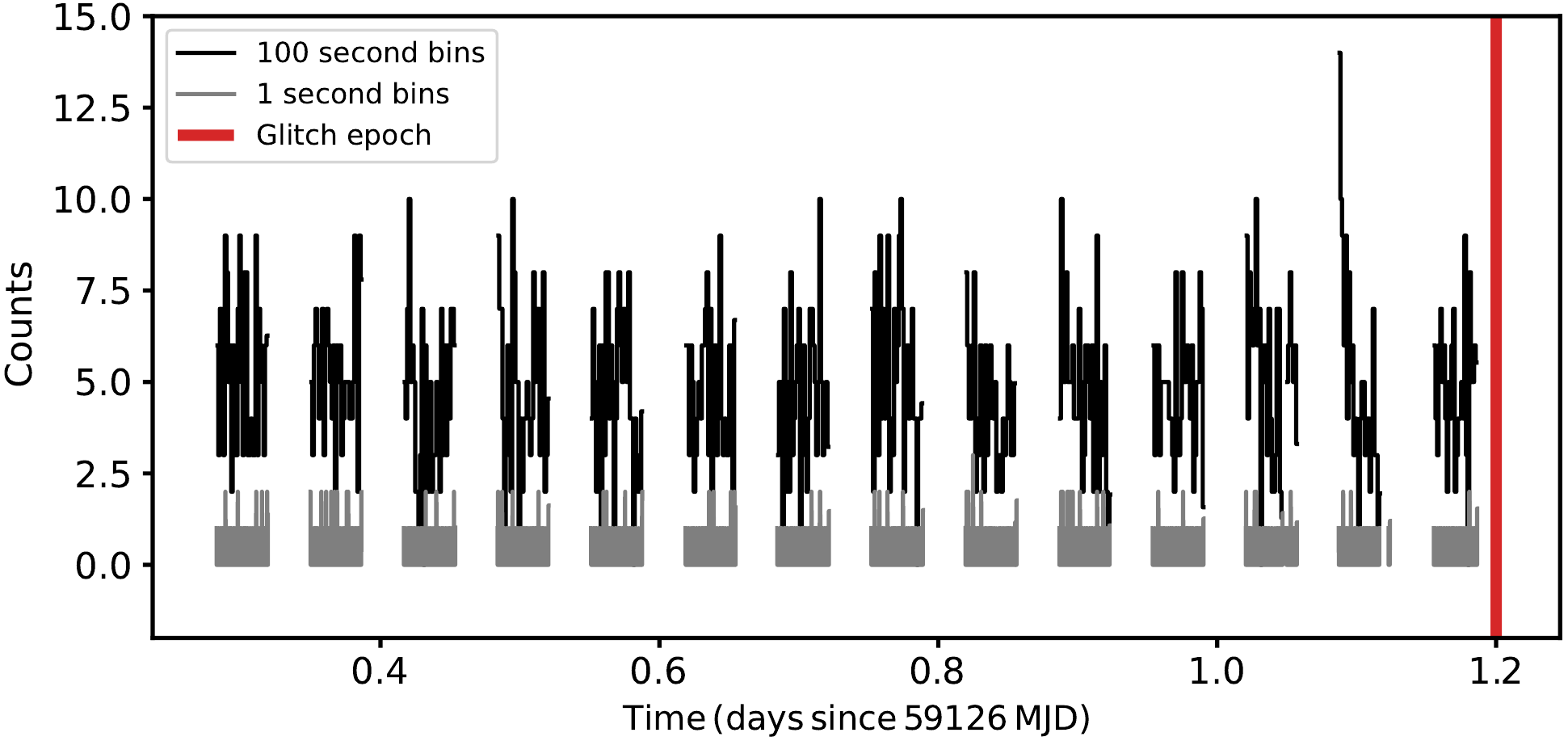}
\spacingNew{1}
\caption{The October 4 \nustar\ light curve of \src\ in the 3-30 keV range. The gray and black histograms display two different binning of the light curve at 1 and 100 second resolution, respectively. The red vertical line denotes the best fit glitch epoch at $59127.2$~MJD. Note that the \nustar\ observation covers the $1\sigma$ lower-limit on the glitch epoch.}
\label{fig:nusLC}
\end{center}
\end{figure*}

The ensemble picture is then that if \teq{\delta t \sim 10}~hours, the radiation would
not be easily observed by large FoV instruments such as {\sl Fermi}-GBM and {\sl Swift}-BAT in X rays, nor by ZTF or PanSTARRS in the optical. Furthermore, assuming that the glitch-epoch coincides with the October 4 \nustar\ observation (Figure~\ref{fig:nusLC} and Table~\ref{specParam}), we set a $3\sigma$ upper-limit on any associated X-ray flux enhancement to be about $10^{-12}$~erg~s$^{-1}$, or a luminosity of about $10^{34}$~erg~s$^{-1}$ at a distance\cite{zhong20ApJ1935} of 10~kpc. This is highly indicative of further enlargement of the open-field line region, or baryonic loading of the wind, both of which lower its radiative efficiency. A long duration radio afterglow might also be expected where the wind deposits its kinetic energy in the circum-magnetar medium, somewhat analogous to that seen\cite{Gaensler-2005-Nature} for the giant flare of SGR 1806-20. Covering the next anti-glitch epoch with daily cadence will provide pivotal clues elucidating on the physical properties of the outflow; mass content, total energetics, radiative efficiency, etc.

{\bf Context concerning the 2020 April 28 FRB and May 24 radio bursts.} 
The spin ephemerides for \src\ around the 2020 April 28 FRB are not well constrained due to sparse data coverage\cite{younes20ApJ1935}, and the one frequency measurement hours prior to the 2020 April 28 FRB had a $1\sigma$ uncertainty of $2.0\times10^{-6}$~Hz. On the other hand, the timing solution surrounding the May radio bursts\cite{kirsten2020} is based on four ToAs\cite{younes20ApJ1935}. Hence, we cannot place meaningful limits on the size of a putative spin-down glitch at the time of either radio event. Interestingly, deep observations with several radio dishes\cite{bailes21mnras} including the FAST radio telescope\cite{lin2020Natur} did not reveal any pulsed radio emission following these bursts. Hence, while our observational result strongly argues that a spin-down glitch is associated with, and likely facilitated, the production of FRB-like bursts and pulsed radio emission, it is unclear whether it is a universal necessity or the most stringent condition for either. For instance, the time of arrival of the two May and three October radio bursts fully covered a single magnetar rotation, which argues against emission from strictly open magnetic field-lines. Hence, it is possible that an outflow along quasi-polar, but closed field lines may still result in the production of an FRB-like burst\cite{Kunihito20ApJL,younes21NatAs}, yet, without causing a strong spin-down glitch or pulsed radio emission. Other factors could come into play, e.g., the state of the magnetar. For instance during the earlier radio bursts, the \src\ was still in outburst, emitting bright gamma-ray bursts and a pulsed hard X-ray component, both of which are not detected in October/November\cite{borghese22MNRAS}. Continued radio and high energy monitoring of all magnetars is essential to drawing a broader picture that could connect the many transient facets that these topical sources display.

\newpage

\newpage

\newpage 

\pagebreak

\newpage 

\newpage

\newpage 
\pagebreak

\newpage 
\pagebreak

\newpage 
\pagebreak

\begin{table}
\rmfamily
\captionsetup{justification=centering}
\caption{Best fit spin parameters for period covering 2020 June 18 to August 6.}
\label{tabTimSol2}
\begin{center}
\resizebox{0.5\textwidth}{!}{
\begin{tabular}{l c c}
\hline
\hline
R.A. (J2000) \T\B & 19:35:41.64 \\
Decl. (J2000) \T\B & 21:54:16.9 \\
Time Scale \T\B & TDB \\
Ephemeris \T\B & DE405\\
Epoch (MJD) \T\B & 59043.0 \\
$\nu$ (Hz) \T\B & 0.30792914(1) \\
$\dot{\nu}$ (Hz s$^{-1}$) \T\B & $-3.29(3)\times10^{-12}$ \\
Valid Range (MJD) \T\B & 59018.0--59067.0 \\
$\chi^2$/dof  & 10/6 \\
RMS residual (ms) \T\B & 153 \\
\hline
\end{tabular}}
\end{center}
\end{table}

\newpage
\pagebreak

\begin{table*}[t]
\caption{\xmm\ and \nustar\ observations and best fit spectral parameters}
\label{specParam}
\begin{center}{
\resizebox{0.95\textwidth}{!}{
\begin{tabular}{c c c c c c c c c}
\hline
\hline
Observation ID   \T\B & Start date & End date & Exposure & $N_{\rm H}$ & $kT$     & $R_{\rm BB}^{2*}$       & $\Gamma$ & $F_{\rm tot}^{**}$ \\
                 \T\B & (MJD) & (MJD) & (ks) &  $10^{22}~$cm$^{-2}$ & (keV) & (km$^2$) & & ($10^{-12}$ erg s$^{-1}$ cm$^{-2}$)\\
\hline
\multicolumn{9}{c}{\xmm\ observations}\\
\hline
0871191301 \T\B & 59123.7 & 59125.1 & 61 &$2.6\pm0.1$ & $0.42\pm0.01$ & $7\pm1$ & $1.3\pm0.2$ & $3.2\pm0.1$ \\
0872390601 \T\B & 59140.2 & 59140.6 & 29 & (L)         & $0.41\pm0.01$ & $8\pm1$ & $1.0\pm0.3$ & $3.2\pm0.1$ \\
0872390701 \T\B & 59166.1 & 59166.6 & 18 & (L)         & $0.39_{-0.01}^{+0.02}$ & $8_{-1}^{+2}$ & $1.5_{-0.5}^{+0.4}$ & $3.1\pm0.2$ \\
\hline
\multicolumn{9}{c}{\nustar\ observations}\\
\hline
80602313008 \T\B & 59126.3 & 59127.2 & 40 & $1.6\pm0.5$ & $0.6\pm0.1$ & $4_{-1}^{+2}$ & $1.0\pm0.1$ & $4.3\pm0.4$ \\
90602332002 \T\B & 59136.9 & 59137.4 & 21 & (L) & $0.35_{-0.04}^{+0.05}$ & $20\pm13$ & $1.2\pm0.1$ & $4.9\pm0.5$ \\
90602332004 \T\B & 59138.9 & 59139.4 & 18 & (L) & $0.36_{-0.05}^{+0.07}$ & $5_{-3}^{+7}$ & $1.0\pm0.1$ & $4.5\pm0.4$ \\
\hline
\hline
\end{tabular}}}
\end{center}
\begin{list}{}{}
\item[{\bf Notes.}]$^{*}$Derived by adopting a 10~kpc distance. $^{**}$\xmm\ and \nustar\ fluxes are derived in the 1-10 and 1-30~keV range, respectively. Listed uncertainties are at the $1\sigma$ level.
\end{list}
\end{table*}



\end{document}